\documentstyle{mn}
\input{psfig}
\newcommand{\etal}{{et al.~}}

\newcommand{\lta}{\la}
\newcommand{\gta}{\ga}

\newcommand{\kmsmpc}{\>{\rm km}\,{\rm s}^{-1}\,{\rm Mpc}^{-1}}
\newcommand{\kms}{\>{\rm km}\,{\rm s}^{-1}}
\newcommand{\pc}{\>{\rm pc}}
\newcommand{\cm}{\>{\rm cm}}

\newcommand{\Msun}{\>{\rm M_{\odot}}}

\newcommand{\apj}{ApJ}
\newcommand{\apjs}{ApJS}
\newcommand{\aj}{AJ}
\newcommand{\mnras}{MNRAS}
\newcommand{\aap}{A\&A}
\newcommand{\aaps}{A\&AS}
\newcommand{\araa}{ARA\&A}
\newcommand{\nat}{Nature}

\newdimen\hssize
\newdimen\hdsize
\begin{document}


\title[The Origin of the Density Distribution of Disk Galaxies]
      {The Origin of the Density Distribution of Disk Galaxies:\\
       A New Problem for the Standard Model of Disk Formation}
\author[Frank C. van den Bosch]
       {Frank C. van den Bosch \\
        Max-Planck Institut f\"ur Astrophysik, Karl Schwarzschild
         Str. 1, Postfach 1317, 85741 Garching, Germany}


\date{}

\pagerange{\pageref{firstpage}--\pageref{lastpage}}
\pubyear{2000}

\maketitle

\label{firstpage}


\begin{abstract}
  We  present new  models  for  the formation  of  disk galaxies  that
  improve upon previous models by following the detailed accretion and
  cooling of  the baryonic mass, and by  using realistic distributions
  of  specific angular  momentum.   Under the  assumption of  detailed
  angular  momentum  conservation the  disks  that  form have  density
  distributions   that  are  more   centrally  concentrated   than  an
  exponential.   We examine  the  influence of  star formation,  bulge
  formation,  and feedback on  the outcome  of the  surface brightness
  distributions of the stars.   Low angular momentum haloes yield disk
  galaxies with a significant bulge  component and with a stellar disk
  that is  close to exponential, in good  agreement with observations.
  High  angular momentum haloes,  on the  other hand,  produce stellar
  disks that are much more  concentrated than an exponential, in clear
  conflict  with  observations.  At  large  radii,  the models  reveal
  distinct truncation radii  in both the stars and  the cold gas.  The
  stellar truncation radii owe to our implementation of star formation
  threshold   densities,   and  are   in   excellent  agreement   with
  observations.  The  truncation radii in the  density distribution of
  the cold  gas reflect the  maximum specific angular momentum  of the
  gas that has  cooled.  We find that these  truncation radii occur at
  HI surface densities of roughly $1 \Msun \pc^{-2}$, in conflict with
  observations.   We  examine  various  modifications to  our  models,
  including feedback, viscosity, and  dark matter haloes with constant
  density cores, but show that the models consistently fail to produce
  bulge-less disks with  exponential surface brightness profiles. This
  signals a new  problem for the standard model  of disk formation: if
  the  baryonic  component of  the  protogalaxies  out  of which  disk
  galaxies  form have the  same angular  momentum distribution  as the
  dark matter, disks are too compact.
\end{abstract}




\section{Introduction}
\label{sec:intro}

In  the current paradigm for  galaxy formation, set  forth by White \&
Rees (1978) and Fall \& Efstathiou (1980),  galaxies are considered to
form  through the  cooling of  baryons  inside dark matter haloes that
grow  by  means of  gravitational    instability and  acquire  angular
momentum  from cosmological torques.   If the cooling baryons conserve
their   specific angular momentum,  disk   galaxies  will form in  the
centres of the  haloes, with scale lengths  that are in good agreement
with observations (e.g., Dalcanton,  Spergel \& Summers 1997;  Mo, Mao
\& White 1998; de Jong \& Lacey 2000).

However,  detailed hydrodynamical simulations  aimed at  investigating
this process  of galaxy formation have  indicated an important problem
for the cold dark matter (CDM)  model.  In CDM cosmologies haloes form
hierarchically by the merging of many lower  mass haloes.  Because the
cooling in dense, low  mass haloes is very  efficient, the  baryons in
these systems have already cooled by the time they merge with the more
massive protogalaxy.  They reach the  centre of the potential well  by
means of  dynamical friction, through  which they  loose a significant
fraction  of  their specific   angular momentum  to  the  dark matter.
Consequently, the disks that form in these simulations are an order of
magnitude too  small (Navarro  \&  Benz 1991;  White \& Navarro  1993;
Navarro \& Steinmetz 1999).  This has become known as the {\it angular
momentum catastrophe} of disk galaxy formation.

Several solutions  to this problem  have been suggested.  Weil, Eke \&
Efstathiou (1998), Dom\'\i nguez-Tenreiro,  Tissera \& S\'aiz  (1998),
Sommer-Larsen, Gelato  \& Vedel (1999) and   Eke, Efstathiou \& Wright
(2000) argued  that  stellar   feedback  and/or ionizing    background
radiation can prevent the cooling  of gas in  these small mass haloes,
therewith considerably reducing   the angular momentum loss (but   see
Navarro  \& Steinmetz  1997).   An alternative  suggestion has been to
alter the power  spectrum of initial  density  fluctuations, either by
invoking an alternative form  of dark matter (Sommer-Larsen  \& Dolgov
2001), or by resorting to a specific model for inflation (Kamionkowski
\& Liddle 2000).  At  the present, it is still  unclear which of these
solutions, if any, is most successful.  What has become clear, however,
is that if cosmological torques are  indeed the main source of angular
momentum on galactic scales, there is very little room for any angular
momentum loss  if we  want  to explain the  sizes  of present day disk
galaxies.

Motivated by  this  understanding, the past couple   of years a  large
number  of studies have presented analytical  models for the formation
of  disk  galaxies that rely   on the assumption  of  detailed angular
momentum conservation.  Kauffmann  (1996) investigated the  properties
of disk galaxies within this framework, and linked it to the evolution
of damped Ly$\alpha$  absorption systems.  Dalcanton \etal  (1997) and
Mo \etal (1998) investigated the  structural properties of disks, with
emphasis on  the variance induced by  the distribution in halo angular
momentum.  Subsequent  studies included  recipes for  bulge formation,
gas viscosity,  star formation and/or   feedback and investigated more
detailed properties   of   these model  disk galaxies,   such  as  the
Tully-Fisher relation, the gas  mass fractions, and  the origin of the
Hubble sequence  (van   den Bosch  1998,  2000;  Jimenez   \etal 1998;
Natarajan 1999;  Heavens \& Jimenez  1999; van  den Bosch \& Dalcanton
2000;  Firmani  \& Avila-Reese  2000;   Avila-Reese  \& Firmani  2000;
Efstathiou 2000;    Zhang  \&   Wyse 2000;  Buchalter,      Jimenez \&
Kamionkowski 2001; Ferguson \& Clarke 2001).

One important shortcoming of  most of these  models, however,  is that
they make the {\it a  priori} assumption that  the cooled gas arranges
itself in an exponential disk.  However, one of the open issues in the
formation of disk galaxies, is to  actually understand why they reveal
a universal surface  brightness distribution.  If disk galaxies indeed
form as  envisioned in our  standard picture, their  resulting density
distribution  is  directly related  to   the specific angular momentum
distribution of the  protogalaxy.  Motivated  by  the  work of  Mestel
(1963) and Crampin \&  Hoyle (1964), Dalcanton  \etal (1997)  made the
assumption that the  protogalaxy has the angular momentum distribution
of a  uniform  sphere in  solid-body   rotation, and  showed that  the
resulting   disks are actually  more   centrally concentrated than  an
exponential.    Firmani  \& Avila-Reese   (2000)  used  more realistic
distributions of mass and angular momentum for the dark matter haloes,
but again  found  similarly concentrated disks.  Bullock  \etal (2000)
determined the distribution of specific angular momentum in CDM haloes
from high resolution  $N$-body  simulations, and again  concluded that
these distributions will  form overly concentrated disks.   Therefore,
if our picture  for disk  formation is  correct,  the question  arises
whether  the subsequent processes of  star  formation, bulge formation
and feedback can  produce exponential stellar  disks in agreement with
observations.

Here we  present new, but similar,   models for the formation  of disk
galaxies. These models, presented in Section~\ref{sec:models}, will be
used  in  forthcoming  papers  to investigate  a  wide  range  of disk
properties  such as colors  and metallicities, gas mass fractions, and
the Tully-Fisher relation.  In this paper, however, we focus solely on
the  density      distributions   of  the    resulting      disks.  In
Section~\ref{sec:results} we investigate the effects of star formation
and bulge formation on the outcome of  the surface brightness profiles
of the disks.  In  Section~~\ref{sec:trunc} we compare the  truncation
radii  predicted  by the  models  with  observations.  A problem  with
reproducing    low surface    brightness   galaxies   is  discussed in
Section~\ref{sec:DCP},   and    we     summarize our     results    in
Section~\ref{sec:discussion}.

\section{The Models}
\label{sec:models}

\subsection{The basic framework}
\label{sec:framework}

The main assumptions that characterize the framework of our models are
the following: (i) dark matter haloes around disk galaxies grow by the
smooth accretion of mass,  (ii) the angular momentum of  protogalaxies
originates from cosmological torques, (iii) in  the absence of cooling
the baryons have the same distribution of mass and angular momentum as
the dark matter, and (iv)  the cooling baryons conserve their specific
angular momentum.

The two main ingredients that determine the formation and evolution of
a disk galaxy, therefore, are (the  evolution of) the mass and angular
momentum of the virialized    object; $M_{\rm vir}(r,z)$ and   $J_{\rm
vir}(r,z)$\footnote{Throughout  this   paper,  $r$   and $z$  refer to
spherical  radius and redshift,   respectively.}.  We characterize the
angular momentum  of  the   protogalaxies by the   dimensionless  spin
parameter $\lambda = J_{\rm vir}  \vert E_{\rm vir} \vert^{1/2} G^{-1}
M_{\rm vir}^{-5/2}$. Here $E_{\rm vir}$ is the  halo's energy, and $G$
is the gravitational constant. We follow Firmani \& Avila-Reese (2000)
and make the   additional  assumptions that  (i)   the  spin parameter
$\lambda$ of  a given  galaxy is constant  with  time, (ii) each  mass
shell  that virializes  is in  solid  body  rotation,  and (iii)   the
rotation axes of all shells are aligned.

None of these assumptions are necessarily accurate.  For instance, the
assumption  of  smooth  mass  accretion seems  inconsistent  with  the
hierarchical merger picture of  structure formation in a CDM universe.
On the  other hand, the fragility  of disks (e.g.,  T\'oth \& Ostriker
1992) suggests that mergers  can not have played a  dominant r\^ole in
establishing  the  main  properties  of disk  galaxies.   Furthermore,
numerical simulations  suggest that too much merging  results in disks
that are too small, as baryons tend to loose their angular momentum to
the dark matter in the process.  The assumptions regarding the angular
momentum  are   also  questionable.   Nevertheless,  as  we   show  in
Section~\ref{sec:angmom}, these  assumptions result in  haloes with an
angular  momentum  profile  that   is  in  good  agreement  with  high
resolution $N$-body  simulations. Finally, we like to  stress that the
main  goal   of  this  study  is   to  explore  how   the  final  disk
characteristics depend on the  various model ingredients. We therefore
adhere  to simple,  parameterized descriptions  of the  mass accretion
history (hereafter  MAH) and angular  momentum distribution.  Although
perhaps not completely realistic, they provide useful insights.

The main outline of the models is as follows.  We set up a radial grid
between $r=0$ and the present  day virial radius  of the model  galaxy
and we follow the formation and  evolution of the  disk galaxy using a
few hundred time steps. We consider  six mass components: dark matter,
hot gas, disk mass (both  in stars and in  cold gas), bulge mass,  and
mass ejected by outflows from the disk.  The dark matter, hot gas, and
bulge mass are assumed to be  distributed in spherical shells, whereas
the disk stars and cold gas are assumed  to be in infinitesimally thin
annuli.  Each time step  we compute the changes  in these various mass
components  in each  radial   bin.  Below   we describe the   detailed
prescriptions used.

\subsection{The evolution of the dark matter component}
\label{sec:mah}

The backbone of the models is the formation  and evolution of the dark
matter haloes, which is determined by the parameters of the background
cosmological  model and  by the power  spectrum $P(k)$  of the initial
density fluctuations. The parameters  of   importance for the   models
presented  here  are the present  day   matter density $\Omega_0$, the
present day density    in   the  form  of  a  cosmological    constant
$\Omega_{\Lambda}$, the Hubble constant $h \equiv H_0/(100 \kmsmpc )$,
the  baryon   density $\Omega_{\rm   bar}$,   and the    normalization
$\sigma_8$ of $P(k)$.

As discussed above, we make  the assumption that protogalaxies accrete
mass smoothly.  Rather than attempting  to  link the actual  accretion
rate to the cosmological  framework,  using for instance the  extended
Press-Schechter formalism to   construct merger histories,  we adopt a
simple parameterization. This  has the advantage  that we can describe
the mass  accretion   history (hereafter   MAH)  by one  or  two  free
parameters.

\begin{figure}
\psfig{figure=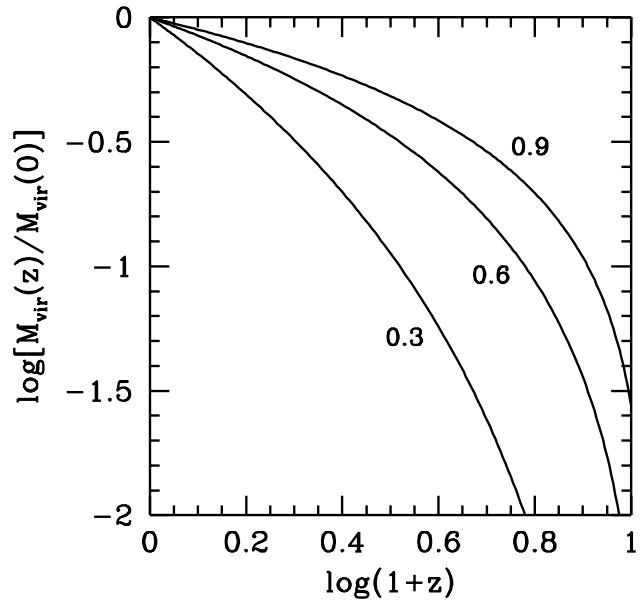,width=\hssize}
\caption{The mass accretion histories (MAHs) defined by
  equation~(\ref{mah}) for $z_f=10$ and three different values of $a$.
  These curves  describe the growth   of halo mass, normalized  to the
  present day  mass,  as function  of   redshift.  A comparison   with
  Figure~1 in Firmani \& Avila-Reese (2000) shows  that these MAHs are
  in reasonable agreement with the $5$, $50$, and $95$ percentile MAHs
  computed using the extended Press-Schechter approximation.}
\label{fig:mah}
\end{figure}

For a given  virialized mass at $z=0$,  $M_{\rm vir}(0)$, we write the
MAH as
\begin{equation}
\label{mah}
M_{\rm vir}(z) = M_{\rm vir}(0) \biggl[ 1 - {\log(1+z) \over
\log(1+z_f)}\biggr ]^{1/a}
\end{equation}
Here $a$  and $z_f$ are free  parameters.  The parameter $a$ describes
whether most of the final  virial mass is accreted early  or late.  In
Figure~\ref{fig:mah}  we plot MAHs for  $z_f = 10$ and three different
values of $a$: $0.3$, $0.6$ and $0.9$.  The corresponding redshifts at
which half the present  day mass is  assembled are $0.57$, $1.26$, and
$2.04$,  respectively.   A  comparison with  MAHs   computed using the
extended  Press-Schechter  approximation  based   on  the  conditional
probabilities     for a  Gaussian       random field  shows  that  the
parameterization~(\ref{mah}) with $z_f=10$  and $0.3 \lta  a \lta 0.9$
comprises $\sim 95$  percent of the typical MAHs  for a typical galaxy
sized  halo, with  $a=0.6$  close to the  average  (cf.  Lacey \& Cole
1993; Firmani \& Avila-Reese 2000; Buchalter \etal 2001).  Clearly, in
reality the  values of $z_f$  and $a$ will depend  on the  mass of the
object. This is not taken into account here.  The aim of this paper is
not to  use   the most accurate MAHs,   but  merely to  use  a  simple
parameterization  to investigate  how    the resulting  disk   surface
densities  depend on  the  MAH.  Therefore,  in what   follows, we use
values of $a$  in the range $0.3$-$0.9$  (with $z_f$ fixed at $10$) to
gauge  the  dependence of our models   on the MAHs. When  required, we
shall explore a wider range of $z_f$ and $a$ values.

The virial  radius of dark  matter haloes  as  function of  redshift is
given by
\begin{eqnarray}
\label{rvir}
r_{\rm vir}(z) & = & 169.0 \, h^{-1} \, {\rm kpc} \, \biggl( {M_{\rm
vir}(z) \over 10^{12} h^{-1} \Msun} \biggr)^{1/3} \biggl({\Omega(z)
\over \Omega_0} \biggr)^{1/3} \nonumber \\
 & & \biggl({\Delta_{\rm vir}(z) \over 178}\biggr)^{-1/3} (1+z)^{-1}
\end{eqnarray}
with $\Delta_{\rm vir}$ the  virial  density, defined as  the  average
density inside $r_{\rm   vir}$  expressed in  terms  of  the  critical
density for  closure. We use  the fitting  formula  of Bryan \& Norman
(1998):
\begin{equation}
\label{deltavir}
\Delta_{\rm vir}(z) = 18 \pi^2 + 82 x - 39 x^2
\end{equation}
with $x=\Omega(z)-1$.  This  fitting formula   is  valid for   a  flat
Universe  ($\Omega_0 +  \Omega_{\Lambda} =   1$)  and accurate  to one
percent in the range $0.1 \leq \Omega(z) \leq 1.0$.

Rather     than attempting to   compute    the density distribution of
virialized  haloes from   extended  secondary infall   models,  as for
instance  in Avila-Reese, Firmani \&  Hern\'andez  (1998), we adopt  a
simple parameterized density  distribution.   We assume that  the dark
matter   virializes such  that at   each  redshift the halo's  density
distribution is given by
\begin{equation}
\label{rhodm}
\rho_{\rm vir}(r) = \rho_s \left( {r \over r_s}\right)^{-\gamma} 
\left( 1 + {r \over r_s}\right)^{\gamma-3}
\end{equation}
with $\rho_s$ and  $r_s$ dependent on $z$  and $M_{\rm vir}(z)$.  This
profile is motivated by both numerical simulations and by the observed
rotation curves of  disk galaxies: numerical simulations of  structure
formation in a  CDM universe suggest  values for $\gamma$ in the range
$1.0-1.5$ (e.g.,   Navarro, Frenk \& White  1996,   1997; Fukushige \&
Makino  1997; Moore   \etal 1998), whereas   rotation curves   of disk
galaxies imply $0 \leq \gamma \lta 1.5$ (van den Bosch \etal 2000; van
den Bosch  \& Swaters 2001). We mainly  focus on models with $\gamma =
1$  for which  equation~(\ref{rhodm})    reduces to the  NFW   profile
(Navarro, Frenk \& White   1997).  However, because of the  relatively
large uncertainties we shall also consider other values of $\gamma$.

The total mass,  energy, and angular momentum of  a halo with  density
distribution~(\ref{rhodm}) are given by:
\begin{equation}
\label{mvir}
M_{\rm vir} = 4 \pi \rho_s r_s^3 f(c),
\end{equation}
\begin{equation}
\label{evir}
E_{\rm vir} = - {1 \over 2} M_{\rm vir} V_{\rm vir}^2 f(c)^{-2} h(c),
\end{equation}
and
\begin{equation}
\label{jvir}
J_{\rm vir} = \sqrt{2} \, \lambda \, r_{\rm vir} M_{\rm vir} V_{\rm vir} f(c) 
h(c)^{-1/2}.
\end{equation}
Here $V_{\rm vir}$ is the circular velocity of the  halo at the virial
radius, $c=r_{\rm vir}/r_s$ is the  halo concentration parameter,  and
$f(x)$ and $h(x)$ are given by
\begin{equation}
\label{fff}
f(x) = \int_{0}^{x} {\rm d}y \, y^{2-\gamma} (1+y)^{\gamma-3},
\end{equation}
and
\begin{equation}
\label{hhh}
h(x) = x \, \int_{0}^{x} {\rm d}y \, f(y) \, y^{1-\gamma} (1+y)^{\gamma-3},
\end{equation}
Equations~(\ref{mah})--(\ref{mvir}) completely specify the dark matter
density distribution at each redshift  once we know  the value of  the
concentration  parameter $c$ as function  of $z$. We  use the model of
Bullock \etal (2001), which gives
\begin{equation}
\label{ccc}
c(M_{\rm vir},z) = C_0 \left( {1 + z_{\rm coll} \over 1 + z} \right)
\end{equation}
with $z_{\rm coll}$ the redshift at which a mass $f \cdot M_{\rm vir}$
collapses.  Here  $C_0$ and   $f$  are constants   that depend on  the
cosmology.

Note that in our approach all haloes at the same redshift and with the
same  mass  have the  same concentration  parameter $c$.   However, in
reality there is a  large spread in  values  of $c$ (e.g.,  Jing 2000;
Bullock \etal 2001).   Furthermore one expects  that $c$ is correlated
with the actual MAHs, such that MAHs with a higher value of $a$ (i.e.,
more   early  accretion)  are   more concentrated (e.g.,  Avila-Reese,
Firmani \& Hern\'andez  1998; Firmani \&  Avila-Reese 2000). We ignore
this  scatter in $c$ and its  associated correlation with $a$. Despite
these     obvious  shortcomings,    the  simplicity    of    the above
parameterizations  allows us to   investigate the main dependencies of
our models, which is our main concern in this paper.

Numerous studies, both analytical and numerical, have shown that the
distribution of the spin parameter $\lambda$ is well described by a
log-normal distribution
\begin{equation}
\label{spindistr}
p(\lambda){\rm d} \lambda = {1 \over \sigma_{\lambda} \sqrt{2 \pi}}
\exp\biggl(- {{\rm ln}^2(\lambda/\bar{\lambda}) \over 2
  \sigma^2_{\lambda}}\biggr) {{\rm d} \lambda \over
  \lambda},
\end{equation}
(e.g., Barnes  \&  Efstathiou 1987;  Ryden 1988;  Cole \& Lacey  1996;
Warren \etal 1992).  In  this paper we gauge  the $\lambda$-dependence
of our models  by constructing models  with $\lambda = 0.028$, $0.06$,
and  $0.129$. These  values correspond  to   the $10$, $50$, and  $90$
percentile  points of  the distribution  of equation~(\ref{spindistr})
with $\bar{\lambda} = 0.06$ and $\sigma_{\lambda} = 0.6$. 

\subsection{Disk formation}
\label{sec:disk}

In order  to compute   the formation and   evolution  of the disks  we
proceed as  follows.   Each time  step $\Delta t$   a  shell with mass
$\Delta M = M_{\rm vir}(t) - M_{\rm vir}(t - \Delta t)$ virializes.  A
fraction $f_{\rm bar} = \Omega_{\rm bar}/\Omega_0$  of this mass is in
baryons, and is heated to the halo's virial temperature
\begin{equation}
\label{tvir}
T_{\rm vir} = {1 \over 2} {\mu m_{\rm p} \over k}  V_{\rm vir}^2
\end{equation}
where $\mu m_{\rm p}$ is the mass per particle, and $k$ is Boltzmann's
constant.  This gas  is added to the  hot gas  component between radii
$r_{\rm  min}$   and $r_{\rm  max}$, whereby we    assume that the gas
follows the same density distribution as the dark matter.  Assuming no
shell crossing we  add $\Delta M$ to the  outer parts of the halo, and
we  thus  set $r_{\rm max}$   equal to the  virial radius  at $t$, and
$r_{\rm min}$  to the radius inside of  which  the total mass (baryons
plus dark  matter) is equal to the  virial mass  at time $t-\Delta t$.
The baryons dissipate energy radiatively,  but are assumed to conserve
their specific angular  momentum.  The time  scale on which  they will
reach centrifugal equilibrium in the disk is given by $t_c \equiv {\rm
max}[t_{\rm  ff}, t_{\rm cool}]$.  Here  $t_{\rm ff}$ is the free-fall
time defined as
\begin{equation}
\label{tff}
t_{\rm ff} = \sqrt{ {3 \pi \over 32 G \bar{\rho}} },
\end{equation}
with $\bar{\rho}$ the average halo density, and
\begin{equation}
\label{tc}
t_{\rm  cool}  =  {3  \over  2}  \mu m_{\rm p}   {k  T_{\rm vir} \over
  \rho_{\rm gas} \Lambda_N(Z_{\rm hot})} {\mu_e^2 \over \mu_e - 1}
\end{equation}
is the cooling   time.  Here $\mu_e$  is  the number of  particles per
electron,  and $\Lambda_N(Z_{\rm  hot})$   is  the  normalized cooling
function for a gas with metallicity  $Z_{\rm hot}$. For $\Lambda_N$ we
use  the   collisional ionization  equilibrium   cooling functions  of
Sutherland  \&  Dopita (1993), assuming  a   Helium mass abundance  of
$0.25$.

Within   our  framework  outlined  in Section~\ref{sec:framework}, the
radius at which the  baryons reach centrifugal equilibrium is computed
as follows.  For a shell with $r_{\rm min} < r < r_{\rm max}$ in solid
body  rotation  with circular  frequency  $\omega_0$ and  with density
distribution $\rho(r)$ the total angular momentum is given by
\begin{equation}
\label{jsh}
J_{\rm shell} = {8 \over 3} \; \pi \; \omega_0 
\int_{r_{\rm min}}^{r_{\rm max}} {\rm d}r \; r^4 \; \rho(r)
\end{equation}
One can also  write  the angular momentum of   the shell as  $\Delta J
\equiv J_{\rm vir}(t) - J_{\rm vir}(t-\Delta t)$, i.e., the difference
in  the  total angular   momentum of the   halo  before and  after the
addition of the new shell of matter. Here $J_{\rm vir}(t)$ is given by
equation~(\ref{jvir}). Thus, for   the shell's circular  frequency one
finds
\begin{equation}
\label{circfreq}
\omega_0 = {3 \over 8 \pi} \; \Delta J \; \left[ 
\int_{r_{\rm min}}^{r_{\rm max}} {\rm d}r \; r^4 \; \rho(r) \right]^{-1}
\end{equation}
At  time $t' =   t + t_c$  we  add these  baryons to  the disk.  Using
detailed conservation of  specific  angular momentum, we compute   the
baryonic mass added to the disk annulus with $r_k < r \leq r_{k+1}$ as
the   baryonic  shell  mass  with cylindrical   radii   $R_k < R  \leq
R_{k+1}$. Here
\begin{equation}
\label{Rcyl}
R_i = \sqrt{r_{i} \, V_c(r_{i},t') \over \omega_0}
\end{equation}
with $V_c(r_{i},t')$   the total circular  velocity  of the  galaxy at
radius $r_{i}$ at  time $t'$.  Throughout  we assume that the  disk is
infinitesimally thin,   and  each  time step    we use   the adiabatic
invariant formalism of Blumenthal \etal (1986) and Flores \etal (1993)
to compute the gravitational contraction of the dark matter induced by
the baryons settling in the disk.

\subsection{Star formation}
\label{sec:sf}

Star formation  in (disk) galaxies  is a complicated  process which is
only  poorly understood.  In particular,   it  is unclear whether,  in
isolated disk galaxies, star formation  is mainly triggered by density
waves (Wyse   1986;  Wyse  \& Silk 1989),    by   local feedback  from
supernovae (Gerola \&   Seiden 1978), or  by cloud-cloud  interactions
(Tan  2000).  Furthermore, although  it is generally accepted that the
star formation  is  self-regulated,  the detailed   interplay  between
gravitational instabilities, supernovae feedback, turbulent viscosity,
magnetic fields, and   star formation is  still  heavily debated  (see
e.g., Firmani \& Tutukov   1994; Silk 1997;  Tan 2000;  and references
therein). Therefore, we use a star  formation recipe that is motivated
by   empirical findings,   so  that despite   our  ignorance, we   are
implicitly taking account of all these detailed physical processes.

Observations have shown that the star formation rates in disk galaxies
are well fit by a simple Schmidt (1959) law:
\begin{equation}
\label{Schmidt_law}
\psi = \varepsilon_{\rm SF} \, \Sigma^{n}_{\rm gas} 
\end{equation}
This  simple empirical law holds over  many orders of magnitude in gas
surface density, and even applies to circum-nuclear starburst regions.
However, when applied  to {\it local}  gas densities,  the Schmidt law
breaks down at large disk radii, where  the star formation is found to
be abruptly  suppressed.  In a  seminal paper, Kennicutt (1989) showed
that these radii correspond to the radii where the gas surface density
falls  below the  critical  surface density given   by Toomre's (1964)
stability criterion:
\begin{equation}
\label{Toomre}
\Sigma_{\rm crit}(R) = {\sigma_{\rm gas} \, \kappa(R) 
\over 3.36  \, G \, Q}.
\end{equation}
Here $Q$ is a dimensionless constant near unity, $\sigma_{\rm gas}$ is
the velocity  dispersion of  the  gas, and  $\kappa$  is the  epicycle
frequency given by
\begin{equation}
\label{epicycle}
\kappa(R) = \sqrt{2} \; {V_c(R) \over R} \left( 1 + {R \over V_c(R)} 
{{\rm d}V_c    \over {\rm d}R} \right)^{1/2}.
\end{equation}
with  $V_c(r)$  the  circular velocity of   the   entire system.   

Solving ${\rm d}\Sigma_{\rm gas}/{\rm d}t = -\psi$, yields
\begin{equation}
\label{sigma_t}
\Sigma_{\rm gas}(t) = 
\left[-{\varepsilon_{\rm  SF} \over m} \Delta t + \Sigma_{\rm
gas}^{1/m}(t-\Delta t) \right]^{m}
\end{equation}
with $m=1/(1-n)$ (cf.  Heavens \& Jimenez  1999).  In each annulus  in
the disk we then compute the mass in stars formed between $t-\Delta t$
and $t$ as
\begin{equation}
\label{mstar}
\Delta M_{*} = A \left[\Sigma_{\rm gas}(t-\Delta t) - 
\tilde{\Sigma}_{\rm gas}(t) \right],
\end{equation}
with $A$  the area of  the annulus and  $\tilde{\Sigma}_{\rm gas}(t) =
{\rm max}[\Sigma_{\rm crit}(t),\Sigma_{\rm  gas}(t)]$.  This way, star
formation  is  not allowed to  deplete  gas to surface densities below
$\Sigma_{\rm crit}$.  

\subsection{Feedback by Supernovae}
\label{sec:fb}

When stars evolve  they put energy  into the interstellar medium (ISM)
which impacts on the further evolution of  the galaxy. By resorting to
an  empirical description  of  the star formation,  we are  implicitly
taking account of the  effects that these  feedback processes have  on
the star formation rate.  What is not taken  into account, however, is
a possible feedback-driven outflow of gas from the  disk.  Here we use
a simple  parametric  model,  similar to  the   ones  used in  various
semi-analytical models for galaxy formation. We assume that the amount
of gas blown out of the disk is proportional to the total energy input
by supernovae (SNe) and  inversely proportional to the escape velocity
squared.  At each time step we compute  the total energy injected into
the ISM at each disk annulus as
\begin{equation}
\label{SN_energy}
E =  \eta_{\rm SN} \, \Delta M_{*} \, E_{\rm SN}
\end{equation}
Here  $\Delta     M_{*}$     is  the      mass  in     stars    formed
(equation~[\ref{mstar}]),  $E_{\rm SN}  =  10^{51}$ergs is  the energy
produced by one SN, and $\eta_{\rm SN}$ is the number of SNe per solar
mass  of stars formed.  We assume  that this energy  drives a galactic
wind, whereby  a  mass $\Delta  M_{\rm  eject}$  is  blown  out of the
halo. Requiring energy balance one obtains
\begin{equation}
\label{mass_eject}
\Delta M_{\rm eject} =  {2 \, \varepsilon_{\rm fb} \, 
\eta_{\rm  SN} \, E_{\rm SN} \over V_{\rm esc}^2} \, \Delta M_{*}
\end{equation}
(cf.  Kauffmann, White  \& Guiderdoni   1993; Natarajan  1999).   Here
$V_{\rm esc}$ is the local escape velocity, and $\varepsilon_{\rm fb}$
is   a  free parameter   that describes what   fraction  of the energy
released by   SNe   is converted  into kinetic   energy   to drive the
outflow. For simplicity,  we assume that the  ejected mass  is forever
lost from the  system: the  ejected  mass is not considered  for later
infall, and  the corresponding  metals  are  not  used to  enrich  the
infalling gas.

\subsection{Bulge formation}
\label{sec:bulge}

Self-gravitating   disks  tend    to  be   unstable    against  global
instabilities such as bar  formation. Here we  follow the  approach of
van  den  Bosch (1998,  2000)  and Avila-Reese  \&  Firmani (2000) and
assume that an unstable disk transforms part of its disk material into
a  bulge component in a self-regulating   fashion  such that the final
disk is marginally stable. 

Motivated by the work of Christodoulou, Shlosman  \& Tohline (1995) we
consider a disk to be unstable if
\begin{equation}
\label{stabalpha}
\alpha_{\rm max} = \max_{0 \leq r \leq r_{\rm vir}} \left(
{V_{\rm disk}(r) \over V_{\rm circ}(r)}\right) < \alpha_{\rm crit}.
\end{equation}
Here  $\alpha_{\rm  crit}$ is a free    parameter, which regulates the
disk-to-bulge ratios  of    the  final model  galaxies,   and  $V_{\rm
disk}(r)$ and  $V_{\rm circ}(r)$  are  the circular velocities  of the
disk (cold gas plus stars)  and the composite disk-bulge-halo  system,
respectively.  

The formation of bulges out of unstable disk material is a complicated
process.  It is   likely  to  involve bars,  which   are efficient  in
transporting gas inwards, and which  subsequently dissolve to form the
bulge. However, the details of the mass flow, and the shape parameters
of the resulting bulge component   are poorly understood, and we   are
forced to make   some ad hoc  assumptions.   If the disk  is  unstable
according to~(\ref{stabalpha}), we transform   cold gas mass from  the
inside  out to the bulge,  whereby we assume  that  the inflow is such
that it does not create a positive  gradient in the surface density of
the cold  gas.  This particular  choice  for extracting bulge material
from the disk  is based on extensive  tests, and is optimized to yield
stellar disks  that are close to exponential.   We shortly address the
influence of these assumptions in Section~\ref{sec:bulform}.

At each time  step we use an  iterative  procedure to compute the  gas
mass transformed to the bulge,  $\Delta M_{\rm bulge}$, such that  the
resulting disk has $\alpha_{\rm max}  = \alpha_{\rm crit}$.  This mass
$\Delta M_{\rm bulge}$ is assumed  to form stars instantaneously  with
100 percent efficiency,  and the SN energy  released by  this burst of
star formation  is added to the  energy released by the quiescent star
formation in the disk at the disk radii from  which the bulge material
originates. This latter  assumption has no important consequences  for
our  results.  If, for  instance, we  were  to deposit  all  SN energy
related with  the bulge  formation at  $r=0$, this  does not alter any
of our conclusions.

We model the bulge as a sphere with a Hernquist density profile:
\begin{equation}
\label{bulgeprof}
\rho_b(r) = {M_b \over 2 \pi} \, {r_b \over r \, (r + r_b)^3},
\end{equation}
where  $r_b$ is a  scale   length (Hernquist 1990).   Given our
poor understanding of the detailed processes involved we use empirical
relations to  compute $r_b$. Andredakis,  Peletier \&  Balcells (1995)
have shown that  the effective  radius $r_e$  (defined  as the  radius
encircling half  of  the projected  light) is  directly related to the
total $B$-band luminosity of the bulge by the empirical relation
\begin{equation}
\label{emprel}
M_B = -19.75 - 2.8 \, {\rm log}(r_e).
\end{equation}
We use this relation to compute the  scale length of the bulge, taking
account of the fact that for a Hernquist  sphere $r_e = 1.8153 \, r_b$
(Hernquist   1990).  In practice, our results    do not depend on this
simple scaling assumption.   The main parameter  for the bulge is  its
total  mass;  changes in its  actual density  distribution are  only a
second-order effect, and do not influence our results.

\subsection{Stellar population modeling \& chemical evolution}
\label{sec:stelpop}

In order  to convert the stellar  masses into luminosities we  use the
latest version  of the  Bruzual  \& Charlot  (1993) stellar population
synthesis models. These models  provide the  luminosities $l_{i}(t,Z)$
of a single burst stellar population with a total mass of $1 \Msun$ as
function of age $t$ and  metallicity $Z$ in various optical  passbands
$i$.  In each disk annulus we keep track of the amount of stars formed
at each time step, $\Delta M_{*}(t_k)$, as  well as the metallicity of
the cold gas $Z(t_k)$.  This allows us to compute the total luminosity
in that annulus, at any time step $t_j$, as
\begin{equation}
\label{sp_lum}
L_{i}(t_j) = \sum_{k=1}^{j} l_{i}[t_j-t_k,Z(t_k)] \, \Delta M_{*}(t_k).
\end{equation}

In order to model  the chemical enrichment  of the  ISM we follow  the
standard instantaneous recycling approximation (IRA). We assume that a
fraction ${\cal  R}$ of the  mass in  stars formed is  instantaneously
returned to the cold gas  phase with a yield  $y$ (which is defined as
the fraction of mass converted into stars that  is returned to the ISM
in the form of newly produced metals). In each disk annulus, and at
each time step, mass conservation  implies
\begin{equation}
\label{mcold}
\Delta M_{\rm cold} = \Delta M_{\rm cool} - (1-{\cal R}) \Delta M_{*} -
\Delta M_{\rm eject}
\end{equation}
and for the mass in metals one thus obtains
\begin{eqnarray}
\label{mmetal}
\Delta M_{\rm metal} & = & Z_{\rm hot} \Delta M_{\rm cool} - 
Z_{\rm cold}  \Delta M_{\rm eject} - \nonumber \\
 & & Z_{\rm  cold} (1-{\cal R}) \Delta M_{*} + y \Delta M_{*}
\end{eqnarray}
We use  these two equations to  track the evolution of the metallicity
of the cold gas in the disk, $Z_{\rm  cold}$, as function of both time
and radius.  Note that  the mass that is  ejected is forever lost from
the galaxy.  The associated metals are not used  to enrich the hot gas
in the halo, and $Z_{\rm hot}$ is therefore constant with time.

\subsection{Detailed description of model parameters}
\label{sec:param}

We distinguish two different sets of parameters.  The first set, which
we  call the {\it galaxy   parameters}, consists of $M_{\rm  vir}(0)$,
$\lambda$, and $a$, and specify  a particular model galaxy. The second
set, the  {\it model parameters},  consists of parameters that specify
the particular formation model. The model parameters are as follows:
\begin{itemize}
\item{} Cosmology: $\Omega_0$, $\Omega_{\Lambda}$, $\Omega_{\rm bar}$,
  $h$,   $\sigma_8$.   In this paper    we  restrict ourselves to  the
  currently   popular $\Lambda$CDM cosmology   with $\Omega_0  = 0.3$,
  $\Omega_{\Lambda}=0.7$, $h=0.7$,   and   $\sigma_8  =   1.0$.  These
  parameters   are currently    favored  by     a  large body       of
  observations. For the baryon  density  we adopt $\Omega_{\rm  bar} =
  0.019  \,  h^{-2}$ as suggested  by   the observations of primordial
  deuterium abundances by Tytler \etal (1999).
\item{} Halo structure: $C_0$, $f$, and $\gamma$. For our $\Lambda$CDM
  cosmology, Bullock   \etal (2001) found  that  dark matter haloes in
  $N$-body simulations have  $\gamma  = 1.0$, $C_0 =   4.0$ and  $f  =
  0.01$. Unless stated otherwise, we shall adopt these values.
\item{}    Star formation: $\varepsilon_{\rm    SF}$,  $n$,  $Q$,  and
$\sigma_{\rm   gas}$.   For a    large sample   of  disk galaxies  and
circum-nuclear     starburst   regions    Kennicutt   (1998)     found
$\varepsilon_{\rm SF} =   2.5 \times 10^{-4}\Msun {\rm   yr}^{-1} {\rm
kpc}^{-2}$ and $n=1.4$,  while it was  shown in Kennicutt (1989)  that
for $\sigma_{\rm  gas} =  6 \kms$ and   $Q = 1.5$  the  star formation
truncation radii are in  good agreement with observations.   Since our
star formation recipe  is based  on empirical  relations, we keep  our
star formation  parameters  fixed at  these observationally determined
values.
\item{} Feedback: $\varepsilon_{\rm fb}$. Since there are no empirical
  constraints on   this   feedback efficiency  parameter we   consider
  $\varepsilon_{\rm fb}$ a free parameter.
\item{} Bulge formation: $\alpha_{\rm crit}$.  This parameter sets the
  amount of self-gravity of the disk above which  we transfer cold gas
  from the disk to  the bulge.  Numerical simulations by Christodoulou
  \etal  (1995) found  $\alpha_{\rm crit}  = 0.7$ for  a gaseous disk,
  which is what we adopt throughout.
\item{} Stellar populations and  chemical enrichment: $\eta_{\rm SN}$,
  ${\cal R}$, $y$, $Z_{\rm  hot}$, and the choice  for an initial mass
  function (IMF). Throughout we adopt  the Scalo (1986) IMF, for which
  $\eta_{\rm SN} = 4 \times 10^{-3} \Msun^{-1}$ and ${\cal R} = 0.25$.
  We  keep the stellar  yield fixed  at $y=0.02$,  which  is a typical
  value used in chemical  evolution models. Unless stated otherwise we
  adopt  $Z_{\rm hot}  =  0.3 Z_{\odot}$, typical of   the hot  gas in
  clusters (Mushotzsky \& Loewenstein 1997).
\end{itemize}

\section{The density distribution of disk galaxies}
\label{sec:results}

\subsection{Cooling properties}
\label{sec:cooling}

We start our investigation into the structural properties of our model
galaxies by computing the density distributions  of the cold gas disks
in the absence of star formation, feedback, and bulge formation.
To that extent we first examine the  disk mass fractions as function of
the various input parameters.
\begin{figure*}
\centerline{\psfig{figure=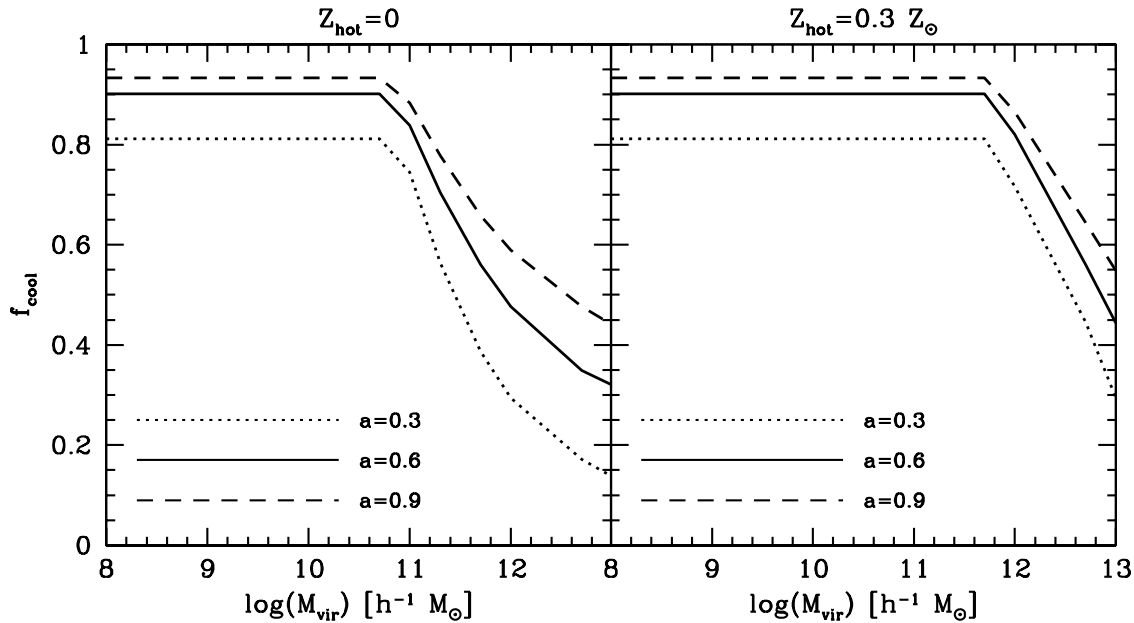,width=15.0truecm}}
\caption{The  fraction, $f_{\rm cool}$,   of baryonic mass inside  the
virial radius that has  cooled  and settled in   present day disks  as
function of  the present day virial mass.  Results are shown for three
different MAHs,  and for two  different metallicities of  the gas. For
low mass systems $f_{\rm cool}$ depends only on the free-fall time and
is independent of halo mass.   For more massive systems $f_{\rm cool}$
is  dominated by the  cooling time scale  and $f_{\rm cool}$ decreases
rapidly with  increasing  mass. Accreting  more mass  at earlier times
(i.e., larger  $a$) and increasing  the metallicity  of  the gas  both
result in larger values of $f_{\rm cool}$.}
\label{fig:allcool}
\end{figure*}

The free-fall time at  each redshift $z$  is the same for all galaxies
as  it depends  only on  the  average  halo  density $\bar{\rho}(z)  =
\Delta_{\rm vir}(z) \, \rho_{\rm  crit}(z)$.  It increases from $0.08$
Gyr at $z=10$ to $2.2$  Gyr at $z=0$.   The cooling time, on the other
hand, depends on the density and  (virial) temperature of the gas, and
is a strong function of $M_{\rm  vir}$. In Figure~\ref{fig:allcool} we
plot the fraction  $f_{\rm cool}$ of  baryons inside the virial radius
that has settled in the  disk at $z=0$ as function  of the present day
virial mass.  Results are shown for three different  values of $a$ and
for both zero-metallicity gas (left  panel) and gas with a metallicity
one-third  Solar (right  panel).  For gas   with  $Z_{\rm hot}=0$  and
$M_{\rm  vir}(0) \lta 10^{11} h^{-1}  \Msun$   one finds that  $t_{\rm
cool} < t_{\rm ff}$ at all redshifts, and between 81 and 93 percent of
the gas has settled  in a disk by $z=0$,  depending  on the MAH.   For
more   massive systems the  cooling  time  exceeds  the free-fall time
during   the later  stages  of  evolution, and   less  mass  can cool.
Consequently,   $f_{\rm cool}$ decreases  strongly  with  mass: for  a
system with $M_{\rm  vir}(0) = 10^{13}  h^{-1} \Msun$ only  between 14
and 44 percent of  the baryons inside the virial  radius can cool.  If
the gas is pre-enriched to $Z_{\rm hot}=0.3 Z_{\odot}$, the cooling is
more  efficient, and $f_{\rm  cool}$  is significantly larger for high
mass  systems.  To summarize,  $f_{\rm cool}$ is  a strong function of
mass, MAH and metallicity,  and one thus  expects strong variations of
the mass fractions of cold gas and stars in galaxies.

\subsection{The density distribution of the gas}
\label{sec:dens}

\begin{figure*}
\psfig{figure=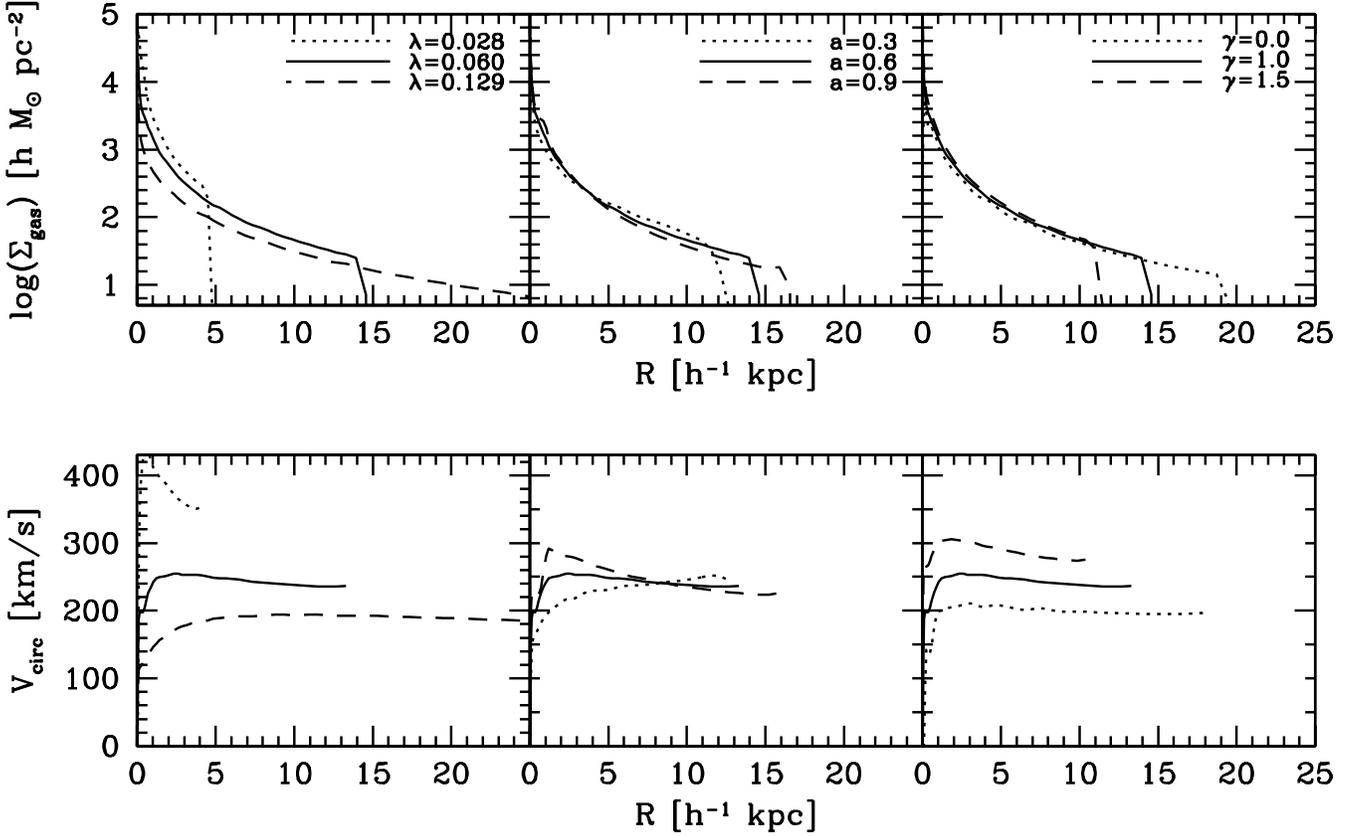,width=\hdsize}
\caption{The upper panels  plot the  surface density distribution   of
cold gas as function of  radius for models  without star formation and
bulge   formation. The lower  panels  plot  the corresponding circular
velocity curves, out to the  cut-off radius of  the disk.  All  models
have the same virial mass of $M_{\rm vir}(0) = 5 \times 10^{11} h^{-1}
\Msun$.  The solid lines correspond  to the model with  $\lambda=0.06$
(median spin parameter), $a=0.6$ (`average MAH') and $\gamma=1.0$ (NFW
halo profile).   The dashed  and  dotted lines show  the  influence of
varying one parameter as indicated  in the top panels  (see text for a
discussion).}
\label{fig:cooling}
\end{figure*}

In Figure~\ref{fig:cooling} we plot  the surface density of the cooled
gas and the total circular velocity at $z=0$ as function of radius (no
star formation or  bulge formation is  included  here).  The different
panels show  the results  of varying one  parameter  with respect to a
fiducial model with $M_{\rm vir}(0) =  5 \times 10^{11} h^{-1} \Msun$,
$\lambda = 0.06$, $a=0.6$, and $\gamma=1.0$  (this model is plotted as
a solid  line in all panels).   In the panels on the  left we vary the
spin  parameter $\lambda$.  As expected,    systems with more  angular
momentum (larger    $\lambda$) result  in  more extended   disks.  For
$\lambda \lta  0.05$  the   compact  disks  that form  are    strongly
self-gravitating,  resulting  in  strongly declining  rotation curves.
Once   bulge formation  is  included,   these systems will   transform
significant  fractions of their disk mass  in a  bulge component, such
that the final rotation curves are in agreement with observations (see
Section~\ref{sec:bulform}).         The    middle    panels         of
Figure~\ref{fig:cooling} plot  the results  for three different  MAHs.
An  earlier  MAH (larger  $a$) yields  disks  that  are more centrally
concentrated, but also more extended.  This owes to the differences in
the cooling histories  of the gas.  Galaxies  with an earlier MAH have
more gas  cooled  at   high  redshifts  when    the halo   is  denser.
Consequently,  that gas  reaches  centrifugal  equilibrium  at smaller
radii.  In addition, the final value  of $f_{\rm cool}$ is higher (cf.
Figure~\ref{fig:allcool}), and this  excess gas cools to  large radii,
resulting in a   more extended disk.   Finally,  in the panels on  the
right we plot  results for three different  values of  $\gamma$.  Less
strongly cusped haloes (lower  $\gamma$) result in more extended disks
and lower values of $V_{\rm circ}$.  This owes to the fact that haloes
with   lower $\gamma$ are    less dense, such    that the  gas reaches
centrifugal equilibrium at larger radii.

As is immediately apparent from  Figure~\ref{fig:cooling}, none of the
disk surface density profiles  resemble an exponential.  In all  cases
the gas is  much more centrally  concentrated, and is well-fitted by a
power-law $\Sigma_{\rm gas}(R) \propto R^{-n}$ with $1 \lta n \lta 2$.
This confirms previous   results by Kauffmann  (1996), Dalcanton \etal
(1997) and Firmani \& Avila-Reese (2000), and  is a direct consequence
of  the detailed conservation  of angular momentum,  and the fact that
the  dark matter haloes have  (broken) power-law density distributions
(see discussions in Seiden, Schulman  \& Elmegreen 1984 and Yoshii  \&
Sommer-Larsen 1989).  In order   to   produce stellar disks with    an
exponential  surface density  distribution out of  these power-law gas
disks one needs  to either (i) have  specifically tuned star formation
efficiencies, (ii) transform part of the central gas disk into a bulge
component, or (iii) somehow remove the central excess  of gas from the
disk.  Below we examine how our particular recipes for star formation,
bulge formation, and feedback fair in this respect.

\subsection{The influence of star formation}
\label{sec:starform}

\begin{figure*}
\psfig{figure=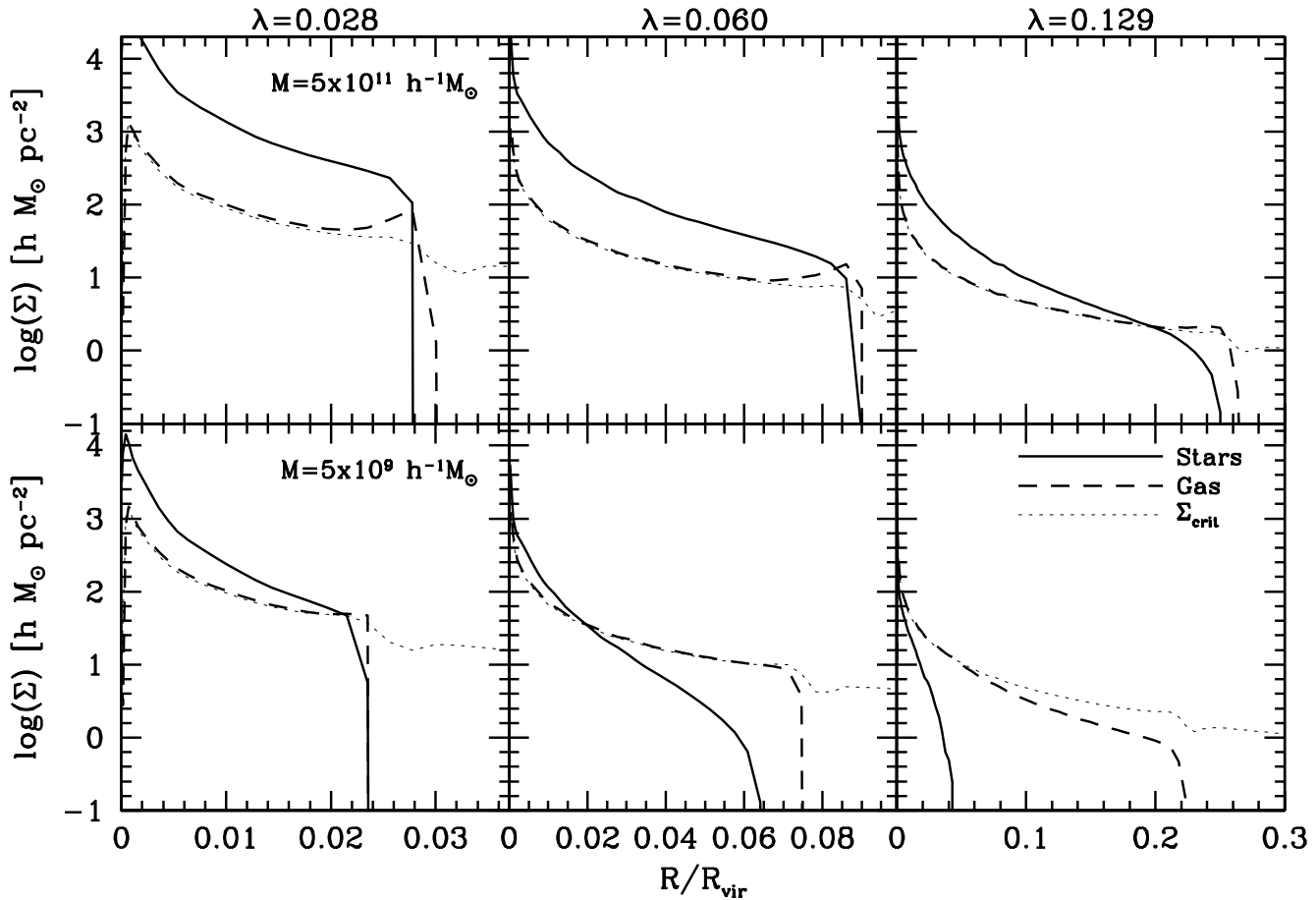,width=\hdsize}
\caption{The $z=0$ surface density profiles of the stars (solid lines)
and cold gas  (dashed lines)  as functions  of the normalized   radius
$R/R_{\rm vir}$. In addition,  the critical surface densities for star
formation are  plotted as thin   dotted lines.  No  bulge formation is
taken into account here.  The six models shown have the same MAH (with
$a=0.6$),  but differ  in  total    mass   and angular   momentum   as
indicated. Except for the model with $M_{\rm vir}(0) = 5 \times 10^{9}
h^{-1}  \Msun$ and  $\lambda =   0.129$, all  stellar disks  are  more
centrally concentrated  than an exponential.  See the  text for a more
detailed discussion.}
\label{fig:sdnobulge}
\end{figure*}

We now   include  star formation  in  our models  using  the empirical
parameters     listed      in        Section~\ref{sec:param}.       In
Figure~\ref{fig:sdnobulge} we  plot the present  day surface densities
of both   the gas  and stars  as  functions  of the normalized  radius
$R/R_{\rm vir}$.  Results are shown for  six model  galaxies that only
differ in total mass and angular momentum  as indicated in the various
panels. 

Except  for the  case with  $M_{\rm  vir}(0) = 5  \times 10^{9} h^{-1}
\Msun$ and $\lambda=0.129$ (lower  right corner) all stellar disks are
more concentrated   than    an  exponential.  Although  the    density
distributions  are   reasonably   close  to  exponential  over   their
intermediate radial range, they harbor a pronounced central cusp and a
distinct cut-off radius.  The surface density of  the cold gas follows
$\Sigma_{\rm crit}$  (indicated by the  thin dotted  lines), except in
the outer regions of  the disk, where the gas  has only  recently been
accreted, and star  formation has not had  sufficient time  to deplete
the density of the gas to $\Sigma_{\rm crit}$.

A    comparison    between   the   upper    and    lower   panels   of
Figure~\ref{fig:sdnobulge} reveals that the lower mass systems produce
disks of lower surface brightness and  with higher gas mass fractions.
As pointed out by previous studies, this  scaling is in good agreement
with observations (e.g., Dalcanton \etal  1997; Mo \etal 1998; van den
Bosch   2000; van      den     Bosch \&   Dalcanton    2000).     From
Figure~\ref{fig:sdnobulge} it is also apparent that lower mass systems
have larger  ratios between the sizes of  the gas disk and the stellar
disk. We return to this issue in Section~\ref{sec:trunc}.

\subsection{The influence of bulge formation}
\label{sec:bulform}

As already eluded to by  Bullock \etal (2000), if  the excess gas mass
in the  centres of  the disks is  transformed  into a bulge component,
this may solve the problem with the overly  concentrated disks. We now
examine   whether the inclusion   of   our particular model  for bulge
formation, based on  the idea  that  self-gravitating disks  transform
part of their material into a bulge, yields exponential stellar disks.
\begin{figure*}
\psfig{figure=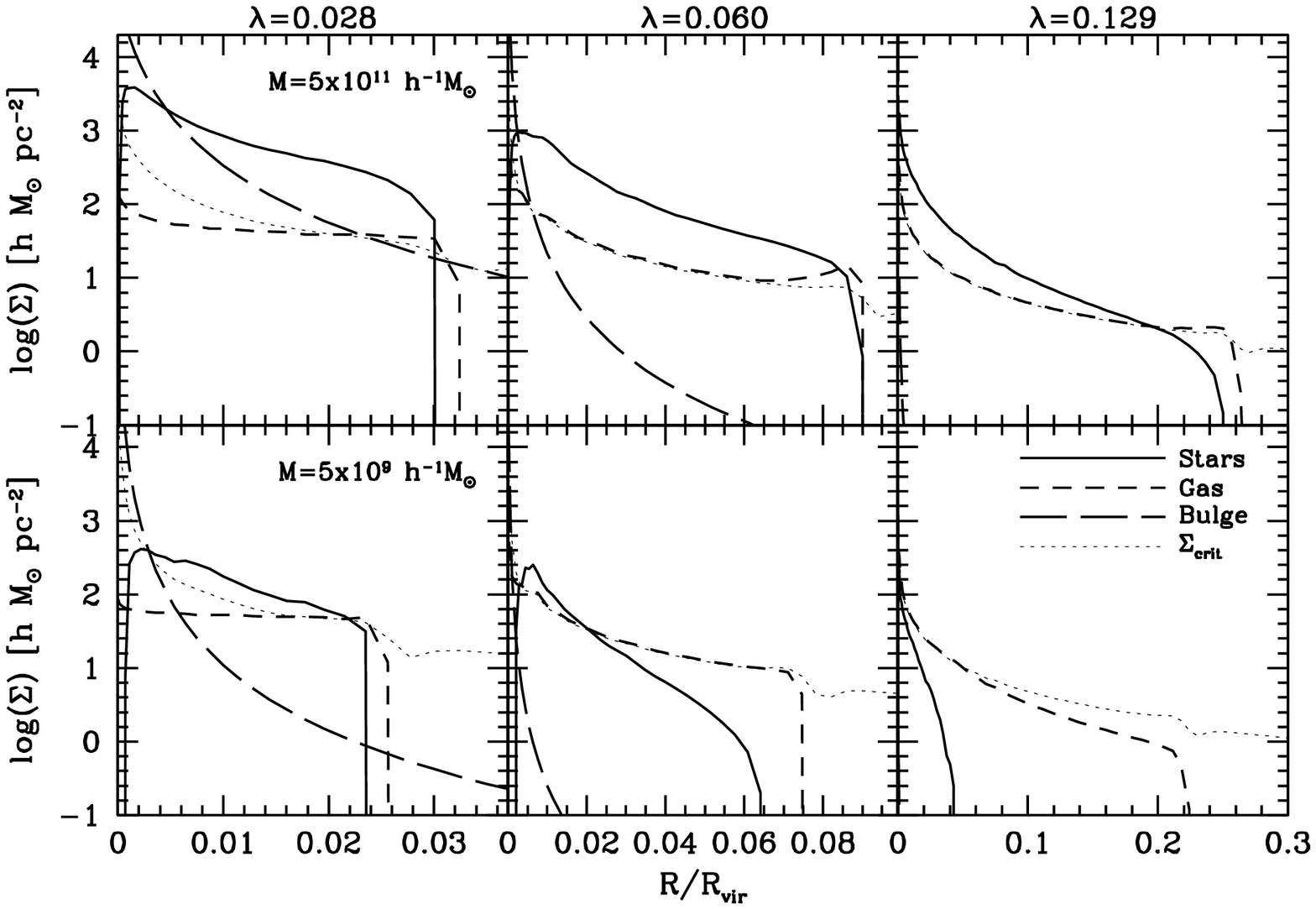,width=\hdsize}
\caption{Same as  Figure~\ref{fig:sdnobulge}, except that now  we have
included   bulge formation.   In  the  cases  with $\lambda=0.028$ and
$\lambda=0.06$ significant bulges have formed out  of  the low angular
momentum disk material,  and the resulting  stellar disks have surface
density  profiles that are   close  to exponential over  their  entire
radial range.  Systems with  high  angular momentum, however,  do  not
produce a significant bulge component, and the stellar disks that form
are   similar       to    the   case       without    bulge  formation
(cf. Figure~\ref{fig:sdnobulge}).}
\label{fig:sd}
\end{figure*}

In Figure~\ref{fig:sd} we plot the $z=0$ surface densities of the same
models as  in  Figure~\ref{fig:sdnobulge}, except  that  we  now  have
included  bulge    formation (using      the  parameters  listed    in
Section~\ref{sec:param}).   The low   angular  momentum  systems  with
$\lambda   =  0.028$   have    formed  massive   bulges  with  $M_{\rm
bulge}/M_{\rm disk} = 0.7 -  0.8$ (i.e., they  are more reminiscent of
S0s than spiral galaxies).  The bulge formation  has depleted the cold
gas in the centre to below $\Sigma_{\rm crit}$, suppressing subsequent
star  formation and  resulting  in a stellar  disk that  is remarkably
exponential.  The  same applies to the  systems with  $\lambda = 0.06$
(middle panels), for  which  $M_{\rm bulge}/M_{\rm disk}  \sim  0.07$,
typical  of late-type Sc spirals.  Clearly,  the  formation of a bulge
component out of the  low angular momentum  disk material seems fairly
successful in  producing exponential stellar  disks.  In the case with
$\lambda=0.129$,  however,  hardly any   bulge  component is  produced
($M_{\rm bulge}/M_{\rm disk} = 0.001$). Consequently, the more massive
models produce, as in the case without  bulge formation, stellar disks
with a central  cusp in   excess  to the outer exponential    profile.
Depending  on the radial  range in  which  the exponential is fit, the
cusp  contains on the order of  20  to 30 percent  of the  mass of the
stellar disk.

Recall that we model  the inflow of the gas  out of which the bulge is
formed   such  that no  positive gradient  is  created  in the surface
density of the cold gas  (Section~\ref{sec:bulge}).  This is a  fairly
ad hoc assumption,  especially since it  is well known that the actual
HI   distribution of disk   galaxies often  reveal  a strong  positive
gradient reflecting a  central hole in the HI distribution\footnote{It
is currently still unclear to what extent these holes are truly devoid
of gas, or whether they are actually filled  with molecular or ionized
gas.}.  However, it  turns out that  the  density distributions of the
stellar   disks   are fairly   insensitive   to  the  details   of the
inflow-model.  If, for  instance,  we model  the gas  inflow such that
{\it all} the  gas  in the  central  bin is transformed into  a  bulge
component before gas  at larger radii  is used (i.e.,  we resort to  a
maximally efficient inflow), the bulge dominated model galaxies reveal
gas disks with large central holes, whereas the stellar disks are only
marginally different  from the  standard inflow model.   Thus, whereas
our  modeling  of the bulge  formation is  admittedly  crude, the main
point is that despite the fact that we have optimized the cooling flow
to produce exponential disks  in systems with relatively large bulges,
we are still left with the problem that the high angular momentum, low
surface  brightness  systems, without  a significant  bulge component,
have  stellar   disks that  are more   centrally concentrated  than an
exponential.
\begin{figure}
\psfig{figure=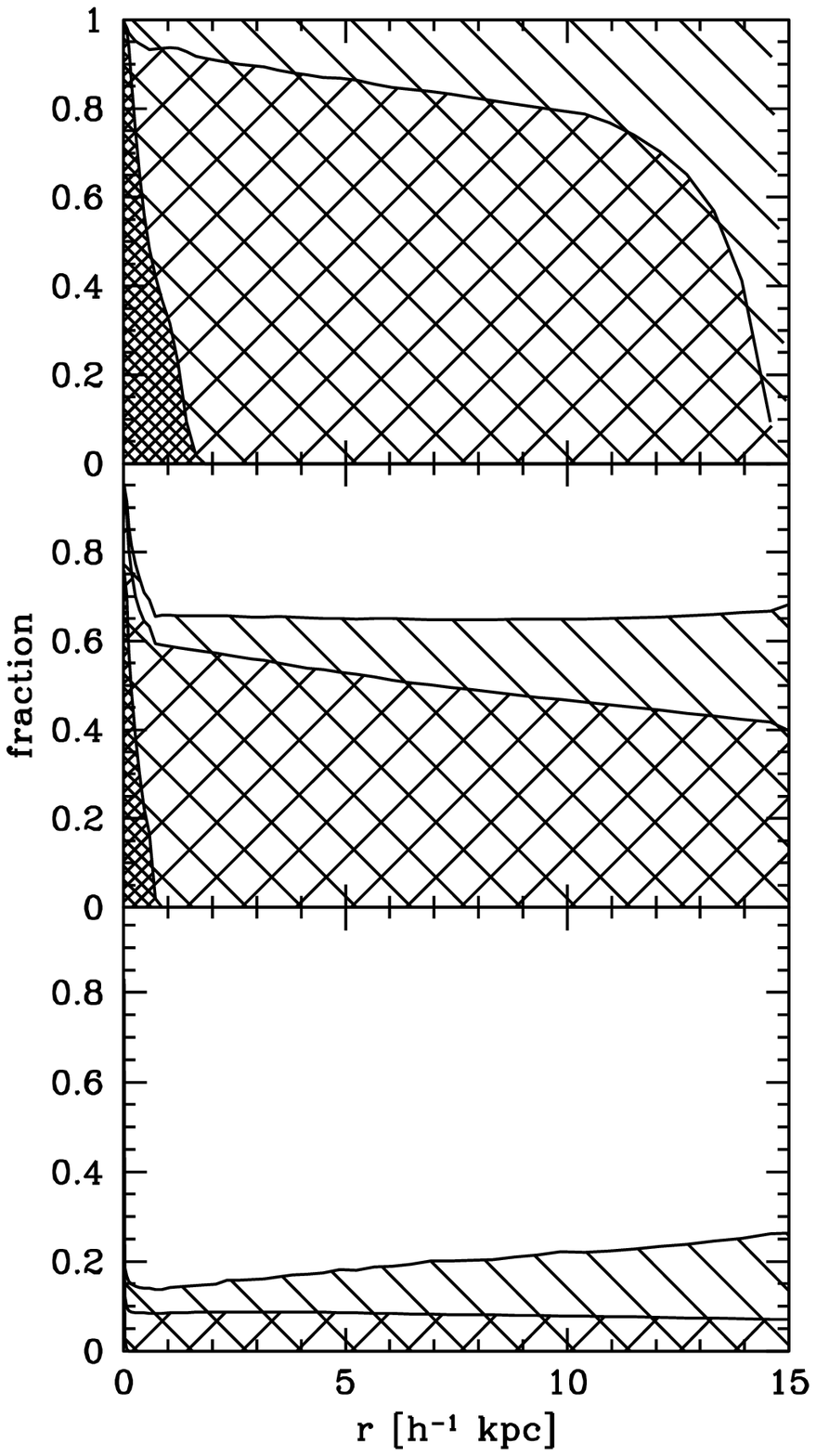,width=\hssize}
\caption{The fractions  of the baryonic  mass in each radial bin that,
at $z=0$, have ended up as bulge mass (densely cross-hatched regions),
disk   stars     (loosely  cross-hatched     regions),    cold     gas
(diagonally-hatched regions), or that have been ejected (white areas).
All three panels correspond to the fiducial  model galaxy with $M_{\rm
vir}(0)=5 \times 10^{11} h^{-1} \Msun$, $\lambda = 0.06$, $a=0.6$, and
$\gamma=1.0$, but differ in the feedback efficiency: $\varepsilon_{\rm
fb} = 0.0$  (upper  panel), $0.02$ (middle   panel), and $0.2$  (lower
panel).  Note how  the baryonic mass  fraction that is ejected depends
only weakly on radius.}
\label{fig:fbfrac}
\end{figure}

\subsection{The influence of feedback}
\label{sec:feedb}

The energy input into the ISM by supernovae can produce galactic winds
that can drive mass out of the disk.   Feedback is therefore a natural
process that can  reshape the density  distributions of the disks.  In
order to  solve the problem with  the excess central densities we need
the feedback  process to be   relatively more efficient in  the centre
than in the outer parts  of the disks.   There are three effects  that
determine these relative  efficiencies   within our simple    feedback
picture: At smaller radii more stars form,  and thus more SN energy is
available  to drive   an   outflow.  At  the   same  time, the  escape
velocities are higher at smaller radii,  suppressing the efficiency of
outflow. Finally, star formation at smaller radii starts earlier, when
the total escape velocity of the system is still  low.  We now examine
how these effects compete and whether or not we can produce bulge-less
disks with exponential surface densities.

In Figure~\ref{fig:fbfrac}  we   plot,  for three   different feedback
efficiencies, the baryonic mass fractions at  $z=0$ in each radial bin
that have ended up as bulge mass,  disk stars, cold  gas, or have been
ejected.   In the case without   feedback (upper panel), virtually all
the gas in  the centre of the  galaxy ends up  in the bulge, while  at
larger radii,  most  of the  gas is  converted  into disk stars.   The
introduction of  a modest  amount of  feedback  with $\varepsilon_{\rm
fb}=0.02$ (middle panel) suppresses both  the bulge formation and  the
formation of disk  stars.   Most remarkably,  the fraction  of the gas
mass  that is ejected is  almost constant with radius.  Apparently the
three effects  mentioned  above cancel  each other  out such  that the
fractional feedback efficiency is virtually independent of radius.  In
the  case with  $\varepsilon_{\rm fb}=0.2$ (lower  panel) the relative
feedback efficiency  is slightly higher  in the centre than  at larger
radii, but the effect is not enough to alleviate  the problem with the
central cusps. We thus conclude that our  simple picture of supernovae
induced feedback does not significantly alter the density distribution
of the resulting  disk galaxies (except for  an absolute  offset), and
thus can not solve the problem of the overly concentrated disks.

\section{Truncation radii}
\label{sec:trunc}

As is evident from   Figure~\ref{fig:sd}, our models   reveal distinct
truncation radii  in the density distributions of  both  the stars and
the  gas. The truncation  radius of the cold  gas reflects the maximum
specific angular  momentum of the baryonic  mass  that has cooled. The
truncation  radius  of the  stars, on  the  other  hand,  reflects the
presence of a star formation threshold density.

In   order to make   a   quantitative  comparison with the    observed
truncation  radii in disk galaxies,  we construct three samples of 100
model galaxies. The  samples only differ in the  value of the feedback
efficiency; all other model     parameters are kept fixed   at   their
fiducial values listed  in Section~\ref{sec:param}.   Masses are drawn
randomly  from a  uniform  distribution with  $3 \times  10^{9} h^{-1}
\Msun  \leq M_{\rm  vir}(0) \leq  3  \times 10^{12} h^{-1} \Msun$, the
spin parameter   is   drawn from  the   probability  distribution   of
equation~(\ref{spindistr})   with   $\bar{\lambda}  =   0.06$      and
$\sigma_{\lambda} = 0.6$,  and the MAH parameter  $a$ is  drawn from a
Gaussian distribution  with $\bar{a}=0.6$ and $\sigma_a  = 0.15$.  For
each model  galaxy we determine the  radii $R_{*}$, $R_{\rm gas}$, and
$R_{\rm  HI}$.  Here we  define  $R_{*}$  as  the maximum radius  with
non-zero surface density of stars, $R_{\rm gas}$ as the maximum radius
with  non-zero surface density  of cold gas,  and  $R_{\rm HI}$ as the
radius at which the surface  density of the cold gas  is equal to $1.3
\Msun  {\rm pc}^{-2}$.  Taking account  of  the Helium mass abundance,
$R_{\rm HI}$ thus  corresponds to  the   radius where the HI   surface
density is equal to $1.0  \Msun {\rm pc}^{-2}$.   In order to  express
these  truncation  radii in terms of  the  scale length of the stellar
disk we fit an exponential to the $I$-band surface brightness profiles
of the models.  Since, as  we have shown above,  not all stellar disks
are  well-fit by an exponential, there  is some  ambiguity involved in
these fits.  After some  experimenting we found adequate  results upon
restricting  the fits  to the radial  interval $0.2  R_{*} \leq R \leq
0.9R_{*}$, which is what we adopt throughout.
\begin{figure*}
\centerline{\psfig{figure=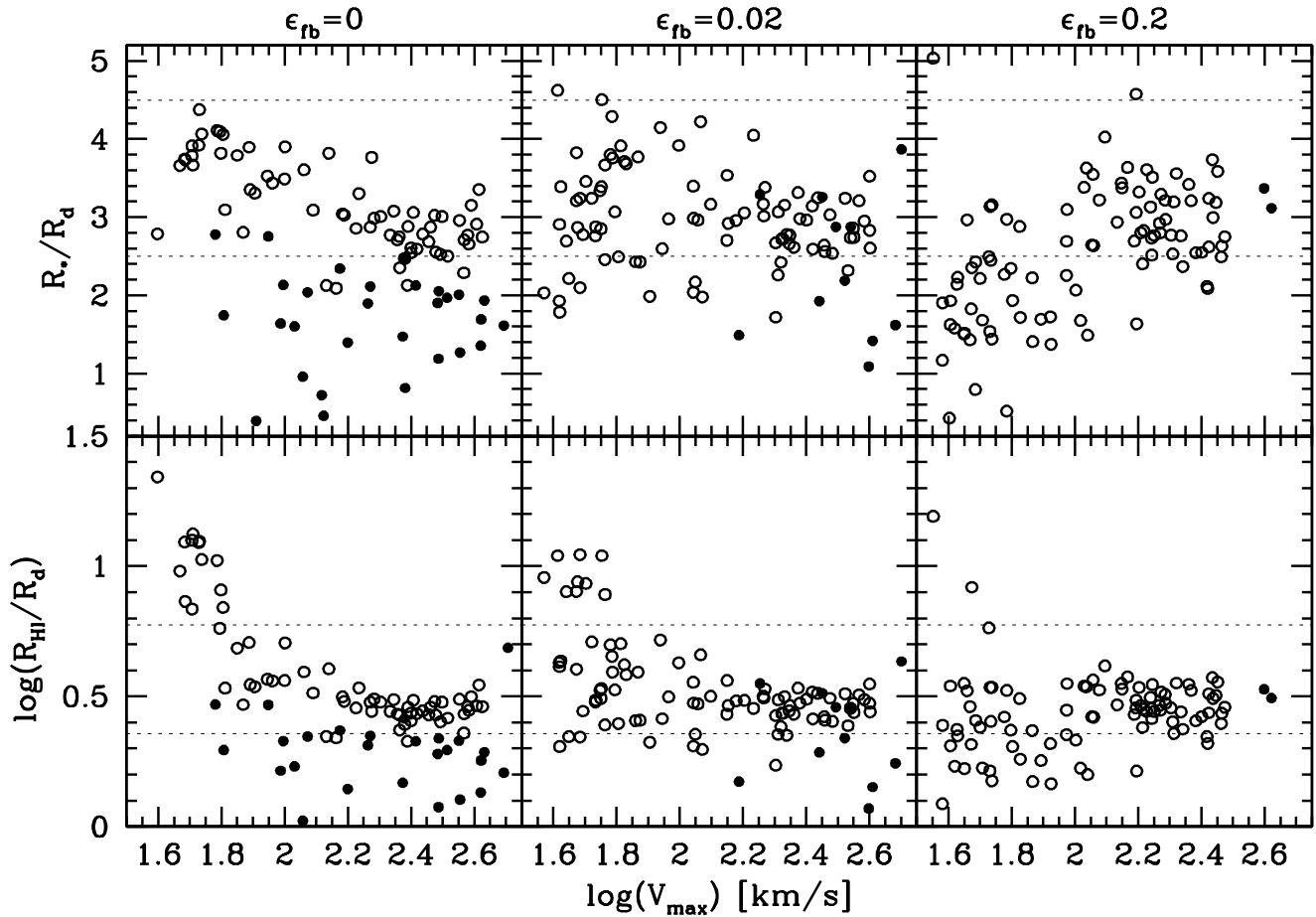,width=\hdsize}}
\caption{The  ratios $R_{*}/R_d$ (upper   panels) and $R_{\rm HI}/R_d$
(lower panels) as functions  of ${\rm log}(V_{\rm max})$.  Results are
plotted for three samples of 100 model galaxies each, that only differ
in the  value  of  the  feedback efficiency as  indicated   above each
column. Open (filled) circles correspond  to models with bulge-to-disk
mass  ratio less    (more) than  $0.2$. The  horizontal   dotted lines
indicate the range  of observed values: $R_{*}/R_d  = 3.5 \pm 1.0$ and
$R_{\rm HI}/R_d = 4.1 \pm 1.8$. See text for a detailed discussion.}
\label{fig:radvmax}
\end{figure*}

\subsection{The extent of stellar disks}
\label{sec:starsize}

\begin{figure*}
\centerline{\psfig{figure=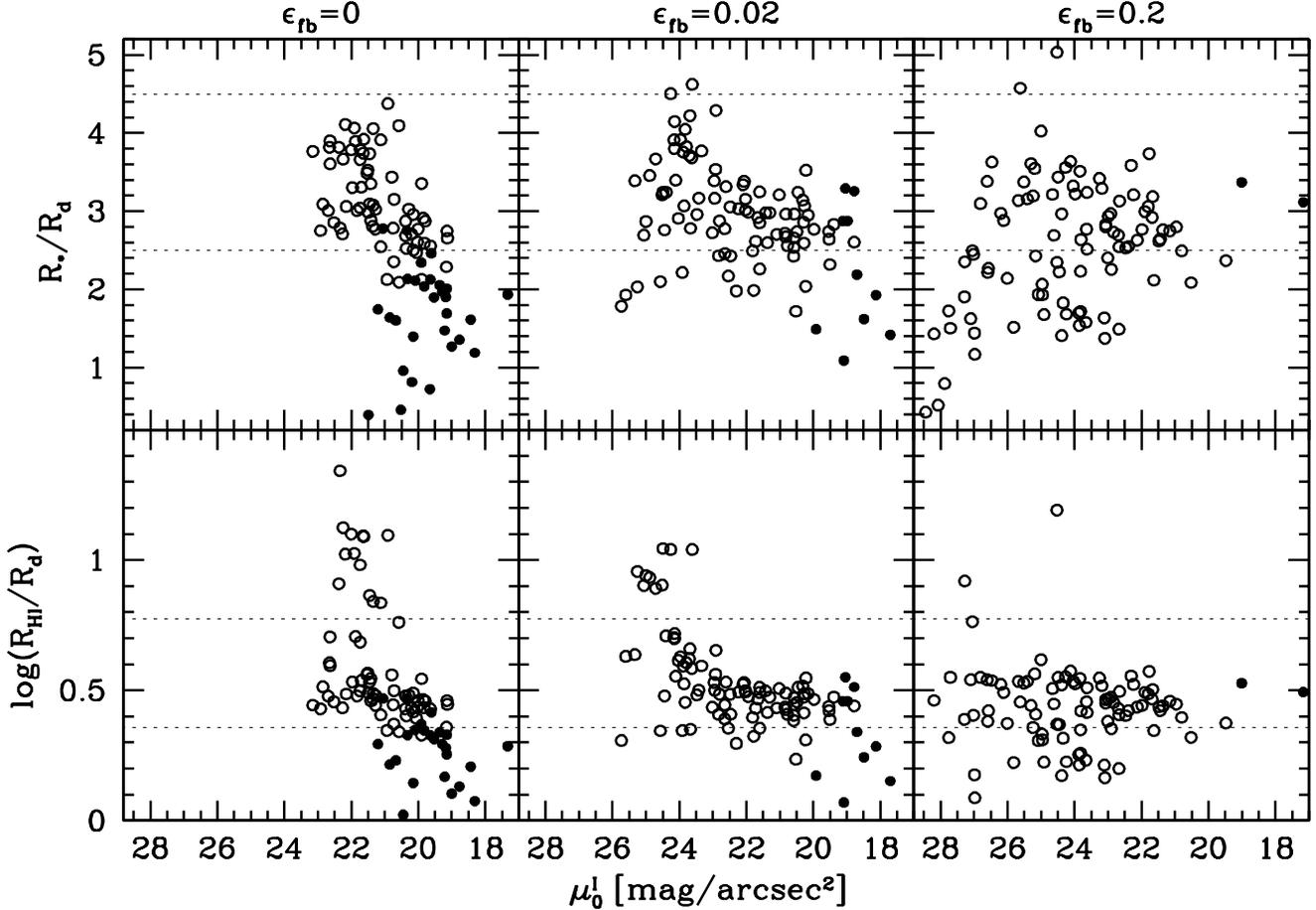,width=\hdsize}}
\caption{Same  as Figure~\ref{fig:radvmax} except that now $R_{*}/R_d$
and $R_{\rm HI}/R_d$ are plotted as functions  of the $I$ band central
surface brightness of the best fit exponential.}
\label{fig:radsb}
\end{figure*}

The upper panels of Figures~\ref{fig:radvmax} and~\ref{fig:radsb} plot
$R_{*}/R_d$ as function of  the maximum of  the rotation curve and the
central $I$-band   surface  brightness of  the   best-fit exponential,
respectively.  Observations  have  shown that stellar  disks in spiral
galaxies reveal truncation radii, $R_{*}$, in the  range from $2.5$ to
$4.5$ disk scale lengths (van der Kruit \& Searle 1981a,b; Romanishin,
Strom \& Strom 1983; Barteldrees  \& Dettmar 1994; Pohlen, Dettmar  \&
L\"utticke 2000; de  Grijs,  Kregel \&   Wesson 2001).  This  range is
indicated by horizontal dotted lines.  It is important to realize that
most studies of  stellar    truncation radii have focussed    on  disk
galaxies with small  bulges.  Therefore,   in  order to  make a   fair
comparison with   our  models, we   plot model galaxies   with $M_{\rm
bulge}/M_{\rm disk} > 0.2$ with separate symbols (filled circles).

In the case  without  feedback (left  panels) the  model galaxies with
$M_{\rm bulge}/M_{\rm disk}    \leq 0.2$ (open  circles) have  stellar
cut-off radii $2 \lta R_{*}/R_d \lta 4.5$, in excellent agreement with
observations.   In   addition,  the  models   predict  that  the  more
bulge-dominated  galaxies  have  $R_{*}/R_d   \lta 2$.  It   should be
relatively straightforward to   test    this prediction   with    deep
photometry of a sample of Sa and S0 galaxies.

Increasing the feedback  efficiency reduces  the average bulge-to-disk
ratio  of  the models and produces   disks with lower  central surface
brightnesses.  In the  case where $\varepsilon_{\rm fb}=0.2$  only two
of   the   model  galaxies have     $M_{\rm   bulge}/M_{\rm  disk}   >
0.2$. Furthermore, the models now predict that low-mass systems should
have values of $R_{*}/R_d \lta 2.5$, significantly  lower than for the
more massive systems. Unfortunately most studies of stellar truncation
radii have focussed on more massive disk galaxies, and a more detailed
comparison of  our models  with observations  has to  await more data,
especially on dwarf and LSB  galaxies.  Nevertheless, it is reassuring
that  our  models with  $M_{\rm   bulge}/M_{\rm disk} \lta  0.2$  have
stellar     truncation radii  in   good   agreement with observations,
supporting   the  idea  that they    originate  from a star  formation
threshold density set by Toomre's stability criterion.

\subsection{The extent of gas disks}
\label{sec:gassize}

Little is known about the truncation radii of the gaseous component of
disk galaxies, mainly due to the fact that the gas is generally probed
by the  HI.  Since at column densities  below  $\sim 10^{19} \cm^{-2}$
one expects  the gas to be ionized   by the cosmic  background flux of
ionizing photons (e.g., Sunyaev  1969; Maloney 1993), a  truncation in
the  HI does not  necessarily reflect the true  outer edge  of the gas
disk.  Therefore, one typically expresses the  size of the gas disk by
the radius $R_{\rm  HI}$ at which  the HI surface  density is equal to
$1.0 \Msun {\rm pc}^{-2}$ ($1.2 \times 10^{20} \cm^{-2}$).

The lower panels of Figures~\ref{fig:radvmax} and~\ref{fig:radsb} plot
$R_{\rm  HI}/R_d$   as  function of    $V_{\rm  max}$  and  $\mu_0^I$,
respectively.   Systems   with $M_{\rm   bulge}/M_{\rm disk}   >  0.2$
typically have  $R_{\rm HI}/R_d \lta 3$.  For  the galaxies with lower
bulge-to-disk  ratios we find that $R_{\rm  HI}/R_d = 2.8 \pm 0.5$ for
$V_{\rm max} > 150 \kms$ and $R_{\rm HI}/R_d \gta  3$ for $V_{\rm max}
< 150 \kms$. For a sample of 73  dwarf galaxies with $V_{\rm max} \lta
100 \kms$ Swaters (1999) found an average of $R_{\rm HI}/R_d = 4.1 \pm
1.8$ (with  $R_d$  measured in the  $R$-band, which  we assume   to be
roughly equal to that  in the $I$-band).   This  is in  good agreement
with  our   models  with  $\epsilon_{\rm fb}=0$     and $\epsilon_{\rm
fb}=0.02$.  However, the model with $\epsilon_{\rm fb}=0.2$ yields and
average of $R_{\rm HI}/R_d = 2.3$ for galaxies with $V_{\rm max} < 100
\kms$, which is only marginally consistent with the data.
\begin{figure*}
\centerline{\psfig{figure=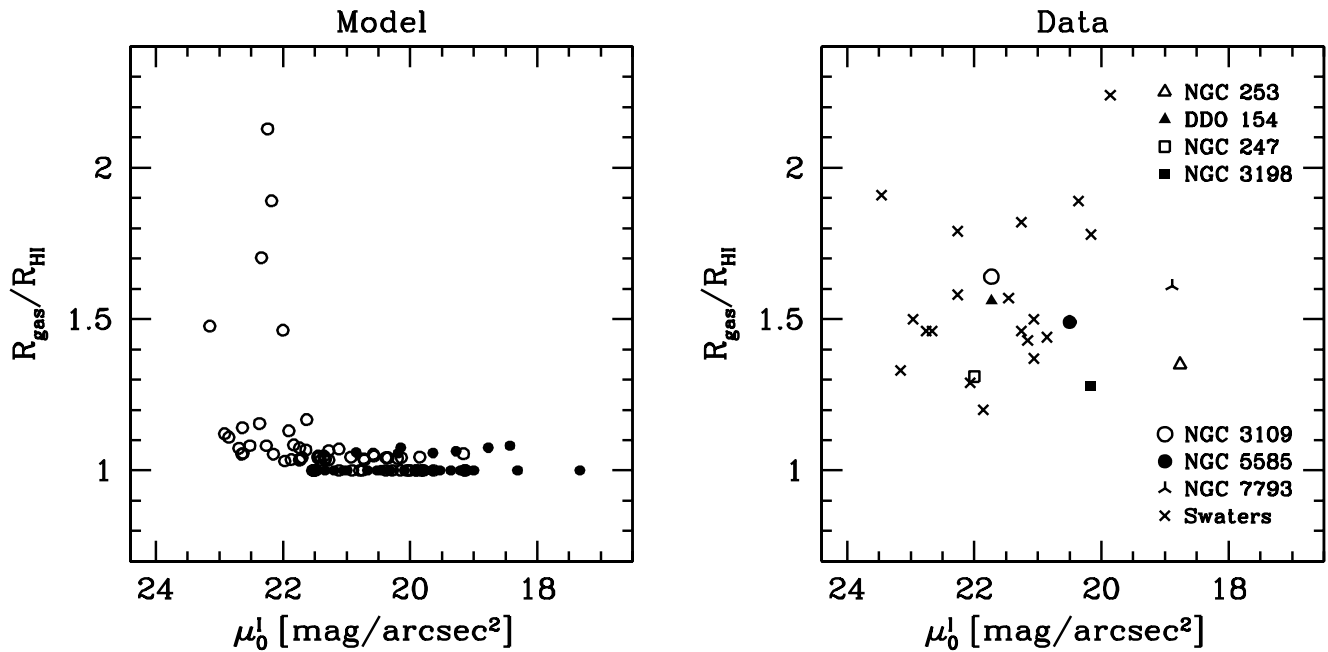,width=\hdsize}}
\caption{The   ratio  $R_{\rm gas}/R_{\rm   HI}$,  between  the actual
cut-off radius  of the gas disk ($R_{\rm  gas}$) and the  radius where
the HI surface density equals  $1.0 \Msun \pc^{-2}$ ($R_{\rm HI}$), as
function of the $I$-band central surface brightness.  The panel on the
left  shows the results  for   a sample of    100 model galaxies  with
$\epsilon_{\rm fb}=0.0$ (all other  model parameters are set to  their
fiducial values,  see   Section~\ref{sec:param}).  Symbols are as   in
Figures~\ref{fig:radvmax} and~\ref{fig:radsb}.  Note how, except for a
few low mass LSB galaxies, all models  have $R_{\rm gas} \simeq R_{\rm
HI}$; the truncation of the model gas disks occurs close to the radius
where $\Sigma_{\rm HI}  =  1.0  \Msun  \pc^{-2}$.  This is   in  clear
conflict with the data for 26  disk galaxies with readily available HI
surface brightness profiles, plotted in  the right panel. Here $R_{\rm
gas}$ is  the  maximum  radius  out to  which  HI  (or  H$\alpha$)  is
detected, and thus  corresponds to a  {\it lower} limit on the  actual
cut-off radius.   Data is taken   from the following sources:  DDO~154
(Carignan \& Beaulieu   1989),  NGC~247  (Carignan \& Puche    1990a),
NGC~3198 (Begeman 1989), NGC~3109  (Jobin \& Carignan  1990), NGC~5585
(C\^ot\'e, Carignan \&   Sancisi (1991), NGC~7793 (Carignan  \&  Puche
1990b), and Swaters (1999). If  no $I$-band surface brightness data is
available we used the colors of de Jong (1996) for conversion. NGC~253
is a  special case.  This  galaxy in  the Sculptor  group  is the only
galaxy in which ionized gas has  been detected at radii beyond $R_{\rm
HI}$ (Bland-Hawthorn, Freeman \& Quinn 1997).  With $R_{\rm HI} = 8.6$
kpc (Puche,  Carignan \&  van  Gorkum 1991)  and  H$\alpha$ and  [NII]
detected out to $11.6$   kpc, this implies   a lower limit  of $R_{\rm
gas}/R_{\rm HI} = 1.35$.}
\label{fig:rHIrgas}
\end{figure*}

In Figure~\ref{fig:rHIrgas} we plot the ratio $R_{\rm gas}/R_{\rm HI}$
as  function of $\mu_{0}^{I}$  for  the model  without feedback  (left
panel).   Models  with $\epsilon_{\rm  fb}=0.02$ and  $0.2$ yield very
similar results.  Except for  low mass LSB galaxies  with $\mu_{0}^{I}
\gta 22$ mag  arcsec$^{-2}$, we find that  $R_{\rm  gas} \simeq R_{\rm
HI}$, indicating that the  physical truncation of  the cold gas occurs
close to the radius where $\Sigma_{\rm HI} = 1.0  \Msun \pc^{-2}$.  We
can compare these  model predictions  to  data by computing  the ratio
$R_{\rm HI,  max}/R_{\rm HI}$, where $R_{\rm HI,  max}$  is defined as
the maximum radius out to which HI is detected.  $R_{\rm HI, max}$ can
thus be considered a {\it lower} limit on  $R_{\rm gas}$.  Results for
a number of disk galaxies with readily available HI surface brightness
profiles are shown in the right panel of Figure~\ref{fig:rHIrgas}.  As
is evident from  Figure~\ref{fig:rHIrgas}, real galaxies have HI disks
that extend to at least $\sim 1.5 R_{\rm  HI}$.  At these radii the HI
surface density typically has   fallen to $\sim 0.1  \Msun  \pc^{-1}$.
Recall that these maximum radii are lower limits to the actual edge of
the gas disk.  Higher sensitivity HI and/or H$\alpha$ observations are
required to determine  the true extent of the  disks.  For the moment,
however, it is  sufficient to realize  that our models, which  predict
truncation   radii  close  to  where   $\Sigma_{\rm HI}  =  1.0  \Msun
\pc^{-2}$, are in clear conflict with the data.

\section{A new problem for the formation of disk galaxies}
\label{sec:DCP}

As shown above   our models predict  that LSB  galaxies reveal surface
brightness profiles  that   are more  centrally  concentrated  than an
exponential. The majority of observed LSB disk galaxies, however, have
surface brightness profiles that are accurately  fit by an exponential
down to the very  centre  (McGaugh \& Bothun 1994;   de Blok, van  der
Hulst, \&  Bothun 1995; Swaters  1999).   An exception are  the giant,
massive LSB disk galaxies, which have surface brightness profiles that
are more  concentrated   than an exponential,  but   this reflects the
presence of a   relatively massive bulge component (Sprayberry   \etal
1995; Pickering \etal 1997).   Our failure to reproduce bulge-less LSB
disks   with exponential surface  brightness    profiles must thus  be
considered a failure of the models.

In addition,  we have shown that  whereas  our implementation  of star
formation threshold densities  yields stellar truncation radii in good
agreement with observations,  the  models predict truncation radii  in
the  gaseous component  at  too high   surface density.  Both of these
problems indicate  that our models  predict disk galaxies that are too
concentrated, and in what follows  we shall refer  to this as the {\it
disk concentration problem} (hereafter DCP).  Below we examine whether
this DCP  might be  an artifact of   the oversimplified nature  of our
models  by discussing how various  modifications of the ingredients or
assumptions of  our model influence   the density distribution of  the
resulting disks.

\subsection{Star formation, galactic winds, and AGNs}
\label{sec:agns}

Although we adhere to empirical relations to  model the star formation
in  the  disk, one  could easily   envision an alternative  model that
results in relatively lower star formation efficiencies in the central
regions.  Although this might produce stellar  disks that more closely
resemble exponentials,    it does  not solve    the  problem  at hand.
Changing the star  formation efficiencies  only changes the   relative
ratio of gas to stars, while leaving the  central {\it mass} densities
intact.  Therefore, such  modifications will  not  alter the  circular
velocity curves, which, at least for the bulge-less LSB galaxies, seem
inconsistent with observations (see Section~\ref{sec:dm} below).

A more  plausible  solution involves  somehow removing the  excess gas
from the centres  of the galaxies, or preventing  it from reaching the
centre in  the first place.  Feedback processes  seem the most natural
means     to    accomplish   this.      However,     as   shown     in
Section~\ref{sec:feedb}, our    simplistic picture  of   SNe   induced
galactic winds is unable to solve the problem at hand.  What is needed
is a feedback process that is relatively more  efficient at the centre
than  at  larger  radii.   It   is  not  inconceivable  that  a   more
sophisticated treatment of stellar  feedback might actually accomplish
that.  For  instance, the presence of  active galactic  nuclei (AGNs),
which have been  ignored in  our models,  might provide  an additional
amount of energy input exactly at the  centre of the galaxies where it
is most required.  However, numerous studies have  shown that the mass
of the AGN's energy source, the massive black hole, is proportional to
the mass of the   bulge in which   the AGN  is embedded  (Kormendy  \&
Richstone 1995; Ferrarese   \&  Merritt 2000; Gebhardt  \etal   2000).
Since the  DCP is most severe for  bulge-less systems, where  one thus
expects no (significant) AGN, it  seems unlikely that the inclusion of
AGN related  feedback processes can solve  the problem with the overly
concentrated LSB  galaxies.  Nevertheless,  because of our   extremely
limited understanding  of   the various  feedback processes  that  are
likely  to  play  a role  in the   formation and evolution   of (disk)
galaxies, it is premature to exclude feedback as  a mechanism that may
solve  the problem  with the central   concentration of the  LSB model
galaxies..   However, it is very unlikely  that feedback processes can
actually   solve  the problem  with the  truncation  radii  of the gas
component, which manifests itself  at  radii beyond the stellar  disk,
and  feedback as a   solution to both  aspects  of the DCP  thus seems
unlikely.

\subsection{Viscosity}
\label{sec:visc}

We have  not included viscous transport in  our  models.  Viscosity is
efficient in  redistributing the disk's angular momentum distribution,
and thus in  modifying the density  distribution of the disk material.
Numerous studies in the past have in fact argued that,  as long as the
viscosity time scale   is similar to   the star formation  time scale,
viscous  transport has the  natural  tendency  to produce  exponential
disks, independent of the initial density distribution of the gas disk
(Lin \& Pringle 1987; Yoshii    \& Sommer-Larsen 1989; Clarke    1989;
Olivier, Blumenthal \&   Primack   1991; Ferguson \&    Clarke  2001).
However, there are two important caveats  here.  First of all, in most
cases the  resulting stellar disk has a  density  distribution that is
only  exponential in the   outer  regions, while  the central  regions
reveal  a  central cusp, reminiscent  of  our disk profiles. Secondly,
virtually all  studies of viscous disks  start with an initial density
distribution that is less  concentrated than an exponential.  However,
we have  shown that  the  standard model for disk   formation predicts
density distributions  that are already  too concentrated, even before
the onset of viscosity.  Finally, it is worth  stressing that the main
mechanism for viscosity is poorly constrained, and in particular, that
there is no detailed theory that links the time  scale of viscosity to
that of star formation, as required.

Most importantly, viscous transport is  oriented inward in the central
regions and  outward  in the outer  regions.  Thus,  whereas viscosity
seems the natural solution for  the problem with the truncation radii,
it  will at the same  time only aggravate the  problem with the overly
concentrated LSB galaxies. We thus do not  consider viscosity a viable
solution to the DCP.

\subsection{The mass accretion histories of dark matter haloes}
\label{sec:mhis}

The DCP outlined  above only applies to LSB  galaxies.  Throughout, it
has been  assumed that LSB  galaxies only distinguish  themselves from
HSB galaxies in  a larger spin parameter.  Several  studies have shown
that this  assumption is  consistent with a  wide variety  of observed
properties of LSB galaxies (e.g., Dalcanton \etal 1997; Mo \etal 1998;
Jimenez  \etal  1998).  One  important  shortcoming  of these  models,
however,  is that  they offer  no explanation  for the  fact  that LSB
galaxies are less strongly  clustered than their HSB counterparts (Mo,
McGaugh \& Bothun 1994). This  has prompted several authors to suggest
that differences in disk surface  density are driven by differences in
the  amplitude of  the original  density fluctuations  from  which the
galaxies arise (e.g., McGaugh \& de  Blok 1998). In this case, HSB and
LSB galaxies are expected to have correlation functions with different
amplitudes.
\begin{figure}
\psfig{figure=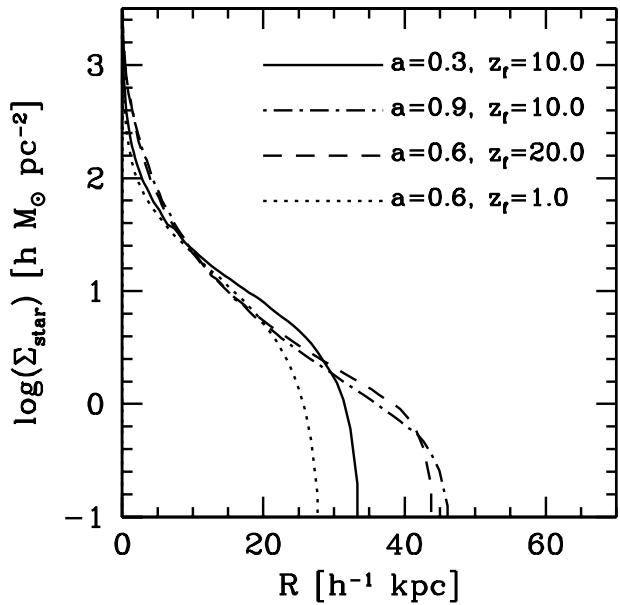,width=\hssize}
\caption{The influence  of the   MAH  on the  stellar surface  density
profiles of the  disks.  Plotted are ${\rm log}(\Sigma_{\rm star}(R))$
for four  model galaxies with  $M_{\rm vir}(0)=5 \times 10^{11} h^{-1}
\Msun$ and $\lambda = 0.129$. The models only differ  in the values of
$z_f$ and $a$ that parameterize the MAH.  Note how despite a wide range
in MAHs, the resulting disks all  are too centrally concentrated to be
consistent with observations.  This illustrates   that the DCP is  not
solved by resorting to specifically tuned mass accretion histories.}
\label{fig:sbmah}
\end{figure}

If indeed  peak height  is the main  parameter that  discriminates LSB
from  HSB galaxies one  expects the  two to  have different  MAHs (the
amplitude  of density perturbations  is directly  related to  the halo
formation epoch).  It is therefore  worthwhile to examine  whether the
disk concentration problem for LSB galaxies may be solved by resorting
to     different    MAHs.     Furthermore,     as    emphasized     in
Section~\ref{sec:mah},  we have  limited ourselves  to a  fairly small
subset of possible  MAHs by only considering models  with $z_f = 10.0$
and  $a$   in  the  range  $[0.3,0.9]$.   Clearly,   a  more  extended
exploration of parameter space is  needed before we can actually claim
a failure of the models to produce realistic LSB galaxies.

In Figure~\ref{fig:sbmah} we  plot  the stellar  surface  densities of
four models with $M_{\rm vir}(0) = 5 \times  10^{11} h^{-1} \Msun$ and
$\lambda = 0.128$ (all other parameters are set to the fiducial values
listed in Section~\ref{sec:param}).   The models only differ  in their
values of $z_f$  and $a$ as indicated. As  is evident,  the MAH mainly
influences the extent of the disk that forms (due to its effect on the
cooling history,  see Section~\ref{sec:cooling}), but has a negligible
effect on the central density of the disk. All four models, which have
wildly   different MAHs, yield   disks that are  inconsistent with the
observed  surface   brightness  profiles of LSB     disk galaxies.  In
particular, the case with  $z_f=1.0$ (dotted line),  which is the most
appropriate for a scenario   in  which LSB   galaxies form  from   low
amplitude density fluctuations,  yields  a surface brightness  profile
that is only approximately exponential over a very small radial range.
Clearly, the DCP can not be solved  by resorting to specifically tuned
MAHs.  Note, however,  that in principle haloes  that form later  will
also be of lower density,  something not taking  into account in these
models  (see discussion in    Section~\ref{sec:mah}). We address  this
influence of  halo density on the surface  brightness  profiles of the
disks in the next section.

\subsection{The nature of dark matter}
\label{sec:dm}

In  the   standard picture for    disk formation, outlined  above, the
density distribution of the  disk is related to  both the density  and
angular  momentum distribution of the  dark matter halo.   If the dark
matter is  cold and  collisionless  it virializes  to produce strongly
concentrated haloes with steep central  cusps.  The disks that form in
these haloes will therefore also be strongly concentrated.  A possible
solution to the  DCP, therefore, might be that  dark matter haloes are
less concentrated,  e.g., they  have a constant  density  core.  It is
noteworthy in this context that numerous studies  in recent years have
claimed that the observed rotation  curves  of dwarf and LSB  galaxies
imply  dark matter  haloes   with constant  density cores  (Flores  \&
Primack 1994; Moore 1994;  McGaugh \& de Blok  1998,  but see van  den
Bosch \etal 2000 and  van den Bosch  \& Swaters 2001). Furthermore, if
LSB galaxies form     from  low amplitude   density  fluctuation  (see
Section~\ref{sec:mhis}), one  expects LSB galaxies  to  be embedded in
low-density dark matter haloes Therefore,  it is worthwhile to explore
to what extent a modification of  the dark matter density distribution
can solve the DCP.
\begin{figure*}
\centerline{\psfig{figure=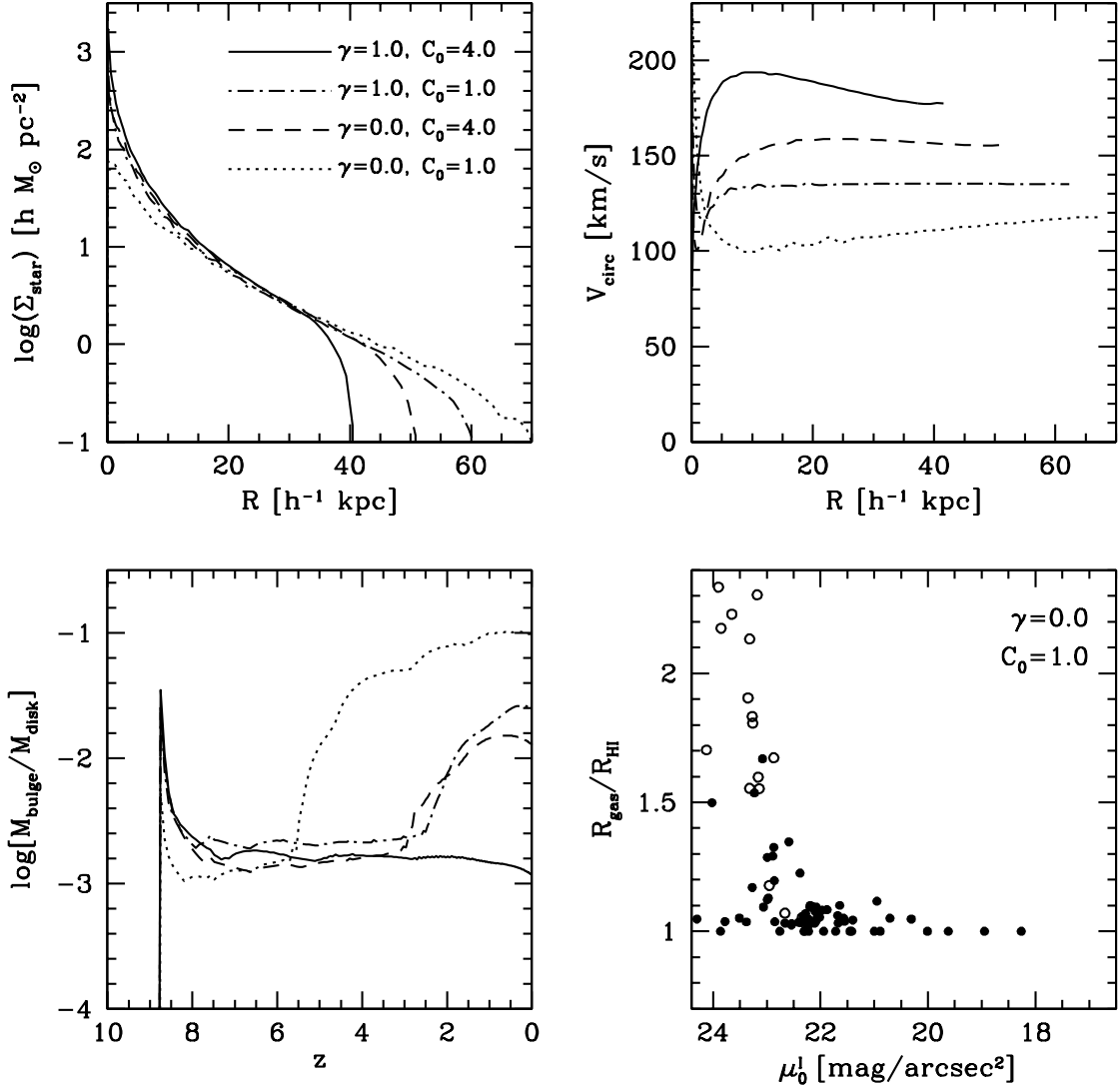,width=15.0truecm}}
\caption{Properties of model  galaxies  with $M_{\rm  vir}(0)=5 \times
10^{11}   h^{-1}  \Msun$, $\lambda =   0.129$   and $a=0.6$  for  four
different dark   matter  density  distributions.    The  solid   lines
correspond  to the   standard $\Lambda$CDM  haloes,  with $\gamma=1.0$
(i.e., haloes    have   NFW density profiles)    and   $C_0=4.0$.  The
dot-dashed   lines correspond to  a   model with a  halo concentration
parameter   that  is four   times smaller  ($C_0=1.0$).   Dashed lines
correspond to a model with a constant density core ($\gamma=0.0$), but
with  the  same concentration parameter   as our fiducial $\Lambda$CDM
model.   The dotted lines,  finally, corresponds to a constant density
cores  with  a core  radius four times  larger  than  for the standard
model.  The upper  panels  plot the surface  densities of  the stellar
disks and  the circular velocity  curves (out to the truncation radius
of the cold gas)  as function of radius.   The lower left panel  plots
the bulge-to-disk ratios (in mass) as function  of redshift.  Note how
less concentrated  dark matter haloes  yield more  extended disks with
close-to-exponential density   distributions and rotation  curves that
rise more slowly, in  better  agreement with observations.    However,
because the disks are more  self-gravitating they also produce  larger
bulges, and the problem of producing bulge-less LSB disk galaxies with
exponential  stellar disks remains.    Finally, the lower right  panel
plots the ratio $R_{\rm gas}/R_{\rm  HI}$ as function of $\mu_{0}^{I}$
for 100  model galaxies with $\gamma=0.0$  and $C_0=1.0$.  Symbols are
as in Figure~\ref{fig:rHIrgas}.  Note how  even for dark matter haloes
with large constant density  cores  the truncation  radii of the   gas
disks in  HSB systems occurs close  to $R_{\rm HI}$, in  conflict with
observations.}
\label{fig:dm}
\end{figure*}

The two  parameters in  our models that   set the dark  matter density
distributions  are the    central    cusp  slope $\gamma$   and    the
normalization  of  the halo concentrations   $C_0$.  In the upper left
panel of Figure~\ref{fig:dm} we plot  the stellar surface densities of
four models with   $M_{\rm vir}(0) = 5  \times  10^{11} h^{-1} \Msun$,
$\lambda  = 0.128$ and  $a=0.6$.  Our standard $\Lambda$CDM model with
$\gamma=1.0$ and $C_0=4.0$   (solid lines) yields a  nearly bulge-free
disk ($M_{\rm bulge}/M_{\rm disk} \simeq  0.001$) with a central  cusp
(cf.    upper right panel   of Figure~\ref{fig:sd}).   The disk  is so
strongly  concentrated  that  the rotation curve   (upper right panel)
reveals a steep  rise followed by a  modest decline.  This is  another
manifestation of  the  DCP, as  the  observed rotation curves  of LSBs
reveal a  slow rise   in the  central  parts  and virtually   never  a
declining part.    The  dot-dashed line shows   the surface brightness
profile for the same model galaxy but with $C_0=1.0$.  This results in
a halo concentration parameter $c$ that is a factor four times smaller
($c=3.7$ compared to $c=14.8$).   Eventhough the mass of the resulting
galaxy is significantly less    concentrated, as is evident  from  the
rotation curves, the surface brightness  of the resulting stellar disk
is only marginally less concentrated.  The main effect of reducing the
halo concentration is to produce a more extended disk, and to increase
the bulge-to-disk ratio  (lower left panel).  Clearly, merely changing
the halo concentrations is not able to solve the DCP.

The dashed lines in Figure~\ref{fig:dm} show  the results for the same
model  galaxy but with $C_0=4.0$   and $\gamma=0.0$, i.e., rather than
changing the halo concentration parameter, we now consider dark matter
haloes with a  constant density core rather than  a $r^{-1}$ cusp. The
results of lowering  $\gamma$ are fairly similar  to those of lowering
$C_0$:  the resulting stellar  disk is somewhat  less concentrated and
more extended than for the case with the NFW dark matter halo, but the
stellar density distribution  is still significantly more concentrated
than  a pure exponential. Furthermore,   the bulge-to-disk ratio is an
order of magnitude larger than in  the standard case with $\gamma=1.0$
and $C_0=4.0$.

The dotted lines, finally, correspond to a model with $\gamma=0.0$ and
$C_0=1.0$. In this case the stellar disk is  close to exponential over
a relatively  large radial range,   and the central  cusp has  largely
disappeared. Furthermore, the resulting rotation curve is still rising
at the  outer   edge  of the  gas  disk,   in   better agreement  with
observations. However, as can  be seen from the  lower left panel, the
bulge-to-disk ratio is,   with $M_{\rm bulge}/M_{\rm disk}=0.1$,   two
orders of  magnitude  larger  than for  the  $\Lambda$CDM  model  with
$\gamma=1.0$ and $C_0=4.0$.   This emphasizes  the robustness  of  the
DCP.  When trying  to lower  the  central densities  of  the  disks by
resorting to  dark  matter haloes  of lower  (central) densities,  the
disks     become  more self-gravitating,   which    results in  larger
disk-to-bulge ratios. The  problem of producing bulge-less exponential
stellar disks therefore remains

Finally, the  lower  right panel  plots  $R_{\rm  gas}/R_{\rm HI}$  as
function of $\mu_{0}^{I}$  for a  sample of  100 model  galaxies  with
$\gamma=0.0$ and $C_0=1.0$.  Compared  to the $\Lambda$CDM  model with
$\gamma=1.0$ and $C_0=4.0$  (left  panel of Figure~\ref{fig:rHIrgas}),
both  the  fraction of low  surface brightness   galaxies  and that of
systems with $M_{\rm bulge}/M_{\rm   disk} \geq 0.2$   have increased.
However, systems  with $\mu_{0}^{I}  \lta 22$ mag  arcsec$^{-2}$ still
have   $R_{\rm gas}/R_{\rm  HI}    \simeq  1$,  in  disagreement  with
observations.  Modifying the structure of  the dark matter haloes  can
thus not  solve  the problem with   the truncation radii  outlined  in
Section~\ref{sec:trunc}.

\subsection{The angular momentum}
\label{sec:angmom}

Under  the assumption  of  angular momentum conservation, the  density
distribution  of the  disk that forms   is a direct  reflection of the
angular momentum distribution of the baryons  in the protogalaxy.  The
DCP might therefore be a consequence of  our specific treatment of the
evolution  of the angular momentum  content.   In particular, we  have
made  four crucial assumptions:  (i) the  specific angular momentum of
the baryons  is conserved, (ii) the  spin parameter of the dark matter
haloes  is constant with time,  (iii) the angular  momentum vectors of
all mass shells are aligned, and (iv) the baryons and dark matter have
the same density and angular momentum distribution

Relaxing assumptions (i) and/or (iii)  in general will only worsen the
DCP.  As discussed in Section~\ref{sec:intro} conservation of specific
angular momentum is required  to produce disks  with the correct  size
distribution.  Furthermore,  gas   with misaligned  angular   momentum
vectors has to align itself in order to form a disk, which it can only
do by  transferring angular  momentum  to the  halo, thus resulting in
even more compact disks.

Since disks  form from the inside  out, one obvious  way of preventing
too much mass from cooling to  small radii is to relax (ii) and assume
that the  spin parameters of protogalaxies  were systematically higher
at  higher  redshift.   However,  although for  each  individual  halo
$\lambda$ is  likely to  vary with redshift,  such a  {\it systematic}
trend  with $z$  seems  to be  ruled  out by  the  fact that  $N$-body
simulations  have  shown  that  the  spin  parameter  distribution  is
virtually independent of mass,  environment and/or redshift (Lemson \&
Kauffmann 1999).  Analytical studies  by Heavens \& Peacock (1988) and
Catelan  \&  Theuns (1996)  have  suggested  a small  anti-correlation
between  the average  spin parameter  and the  peak height  of density
fluctuations.  However,  the amplitude  of  this  correlation is  much
smaller than  the spread in $\lambda$  at any given value  of the peak
height,  and  this  effect  will  thus not  significantly  affect  our
results.
\begin{figure}
\psfig{figure=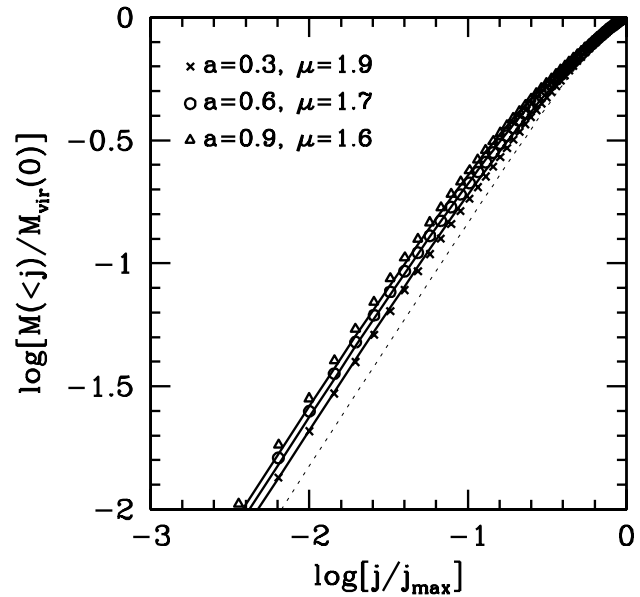,width=\hssize}
\caption{The mass distribution of specific angular momentum for models
  with   $M_{\rm vir}(0)    =  5  \times  10^{11} h^{-1}   \Msun$  and
  $\lambda=0.06$. Results are  shown for three different MAHs: $a=0.3$
  (crosses), $a=0.6$   (open circles), and  $a=0.9$  (open triangles).
  The    solid   lines are      the  best   fit     profiles of    the
  form~(\ref{angmomprof}).  Shown  for comparison is  the distribution
  for a uniform sphere in solid body rotation (thin dotted curve). The
  best fit values  for $\mu$ are listed  in the figure, and vary  from
  $\mu = 1.6$  ($a=0.9$) to $\mu=1.9$ ($a=0.3$),  well in the range of
  $\mu$   values found  by  B00.  This  implies  that  our assumptions
  regarding the evolution of  the angular momentum result in realistic
  angular momentum profiles.}
\label{fig:angmomprof}
\end{figure}

We can further check the validity of  assumption (ii) by comparing the
resulting angular momentum distributions of the  dark matter haloes to
$N$-body  simulations of  structure   formation.  In a  recent  paper,
Bullock \etal (2000, hereafter B00) have shown that dark matter haloes
in a  high  resolution $\Lambda$CDM $N$-body simulation  have specific
angular momentum profiles that are well fit by
\begin{equation}
\label{angmomprof}
m(j) \equiv {M(<j) \over M_{\rm vir}} = 
\mu {(j/j_{\rm max}) \over (j/j_{\rm max}) + \mu - 1}
\end{equation}
Here $M(<j)$ is the halo mass with specific angular momentum less than
$j$, $M_{\rm  vir}$  is  the halo's  virial  mass,  $\mu$  is  a  free
parameter, and $j_{\rm max}$ is the  maximum specific angular momentum
in  the halo.  B00  have shown that the angular  momentum content of a
dark matter  halo is well described  by the pair  $(\lambda,\mu)$, and
that for 90 percent  of the haloes  $1.06 < \mu < 2.0$  with a mean of
$\langle \mu \rangle =  1.25$. We have computed  $m(j)$ for several of
our models.  The  results for  three models with  $M_{\rm  vir}(0) = 5
\times 10^{11} h^{-1} \Msun$ are shown in Figure~\ref{fig:angmomprof},
together with the  best  fit profiles of the  form~(\ref{angmomprof}).
As can be seen, the mass distribution of  specific angular momentum in
our models is  well described by equation~(\ref{angmomprof}) with best
fit  values for $\mu$ in  the range $1.6  \lta \mu \lta  1.9$, in good
agreement with the results of B00.  We have  also tested whether $\mu$
is correlated with  mass, and find a very  weak decrease of $\mu$ with
halo mass\footnote{for  $a=0.6$   we find  that  $\mu$  decreases from
$1.76$ for   $M_{\rm  vir}(0) = 5   \times   10^{9} h^{-1}  \Msun$  to
$\mu=1.68$ for  $M_{\rm vir}(0)  =  5 \times 10^{12}   h^{-1} \Msun$},
again in good agreement with B00.
\begin{figure}
\psfig{figure=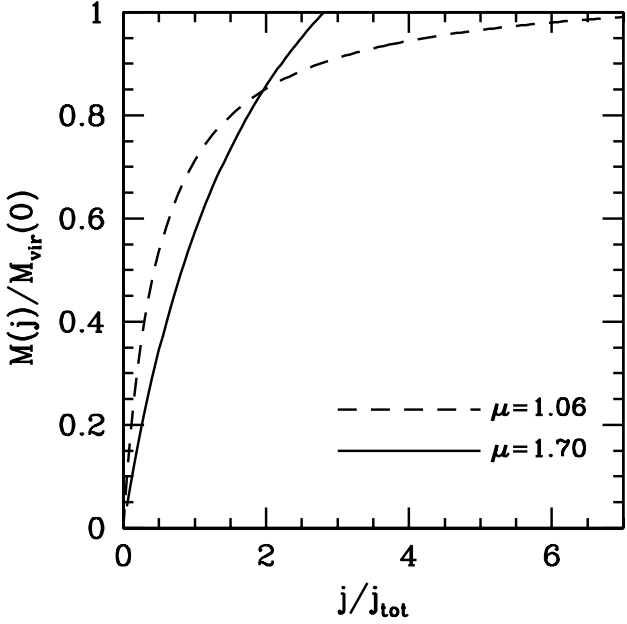,width=\hssize}
\caption{Mass fractions $M(j)/M_{\rm vir}(0)$ as function of $j/j_{\rm
tot}$ for  two different values of  $\mu$.  Decreasing $\mu$ increases
$j_{\rm max}$, and thus the amount of  high angular momentum material,
and    simultaneously increases the   amount   of low angular momentum
material. Lowering $\mu$ thus has the same effect as viscosity.  }
\label{fig:jtot}
\end{figure}

Although our best-fit values  of $\mu$ are  well  inside the  range of
$\mu$   values found by   B00, the spread  we obtain  is much smaller.
However, this is  most likely a reflection of  the fact that our model
haloes have  smooth,    spherical density  distributions  without  any
substructure.  This also   explains why B00  find  a weak  correlation
between $\lambda$ and $\mu$, whereas in our models the $m(j)$ profiles
are    independent of  the  value     of  the  halo  spin   parameter.
Nevertheless,  it is  worthwhile  to examine  whether  the DCP  can be
solved by resorting to haloes with other values of $\mu$.  In order to
address this issue it is useful to rewrite equation~(\ref{angmomprof})
in the form
\begin{equation}
\label{mjprof}
m(j) = \mu { (j/j_{\rm tot}) \over (j/j_{\rm tot}) + (\mu - 1) (j_{\rm
max}/j_{\rm tot})}
\end{equation}
Here 
\begin{equation}
\label{jtot}
j_{\rm tot} = {J_{\rm vir} \over M_{\rm vir}} = j_{\rm max}
\left( 1 - \mu \left[ 1 - (\mu - 1) {\rm ln} \left(
{\mu \over \mu - 1}\right) \right] \right)
\end{equation}
is the  total  specific angular  momentum  of the halo.   Substituting
equation~(\ref{jtot})  in~(\ref{mjprof}) one   can compute  $m(j)$  as
function of $j/j_{\rm  tot}$ for a  given value of $\mu$.  Results for
$\mu =  1.7$ (the average value found  for our haloes)  and $\mu=1.06$
(the 5 percent lower bound of the distribution found by B00) are shown
in Figure~\ref{fig:jtot}.  As  is immediately evident, lower values of
$\mu$  result  in higher values  of $j_{\rm  max}$,   and thus in more
extended disks. However, at the same  time, decreasing $\mu$ increases
$m(j)$  at low values of $j/j_{\rm  tot}$, which implies that the disk
will become more centrally  concentrated, thus aggravating the DCP. In
fact,  lowering $\mu$  mimics the  effects of  viscosity (cf. Zhang \&
Wyse 2000).

It  thus becomes clear that in  order to solve the  DCP it is required
that   the  baryons have an    angular  momentum distribution  that is
completely  decoupled from that of   the dark matter (i.e., assumption
(iv) is in error).  The baryons need  an angular momentum distribution
with  both less low  angular momentum  material  and more high angular
momentum material (i.e., a tail that extends to higher $j_{\rm max}$).
From the  above it is  clear that this  can not  be realized by merely
changing  $\mu$.    Rather,  the  baryons   need an  angular  momentum
distribution  that     is  completely   different  from      that   of
equation~[\ref{angmomprof}]  (see also   van  den  Bosch, Burkert   \&
Swaters 2001).

\section{Discussion \& Conclusions}
\label{sec:discussion}

In our current  paradigm of galaxy  formation, the angular momentum of
protogalaxies originates    from   tidal torques of    nearby  density
perturbations.  It has become  clear that this mechanism provides just
enough  angular momentum to produce   disk galaxies of the right  size
(Fall \& Efstathiou   1980;  Dalcanton \etal  1997;  Mo  \etal 19998),
implying that the baryons can not lose a significant fraction of their
specific  angular momentum. In the standard  picture of disk formation
it    is therefore assumed that   the  baryons conserve their specific
angular momentum when cooling to form the disk. Under these conditions
one can directly  compute  the density distribution  of  the resulting
disk once the actual distribution  of specific angular momentum of the
protogalaxy is known.

Recently, Bullock \etal (2000)  have  determined the specific  angular
momentum distribution of  a large  set   of dark  matter haloes in   a
$\Lambda$CDM  universe, and have shown  that if baryons  have the same
angular momentum distribution    as dark matter,  the   resulting disk
galaxies are  more centrally concentrated  than  an exponential.  This
then raises the question whether or  not baryonic processes related to
star formation, bulge  formation, and   feedback can produce   stellar
disks  with exponential surface  brightness  distributions out of such
highly concentrated gas  disks.  In this  paper we have presented  new
models  for the formation of  disk galaxies to  address this question.
We   use a  simple parameterized   description  of the  mass accretion
histories   of  the  dark matter    haloes,  and, following Firmani \&
Avila-Reese (2000) and  Avila-Reese \& Firmani (2000),  the assumption
is made that the halo  spin parameter is constant  with time.  We have
shown that this implies  specific angular momentum distributions  that
are in good agreement with those obtained by Bullock \etal (2000) from
high resolution $N$-body simulations.  Including cooling and adiabatic
contraction, we have confirmed  the results of Dalcanton \etal (1997),
Firmani    \& Avila-Reese (2000), and  Bullock   \etal (2000) that the
resulting  gas  disks   are  more   centrally  concentrated  than   an
exponential.   Next we included  simple prescriptions to describe star
formation, bulge formation,   feedback,  and chemical evolution,   and
investigated  whether   these  processes  can  transform   the  highly
concentrated gas disks into   stellar disks with close  to exponential
surface brightness distributions as observed.

At first sight our models are  remarkably successful in producing disk
galaxies   with density   distributions  as observed.   The  essential
ingredient is bulge formation, which prevents the formation of systems
with strongly declining rotation  curves.  Furthermore, the process of
bulge  formation  results in stellar disks  with  close to exponential
surface brightness profiles.  This result is fairly insensitive to the
details of the bulge formation process, as long  as it is coupled to a
stability criterion for the    disk (cf.  van den  Bosch   1998).  The
introduction  of a star formation  threshold density,  as motivated by
the results of Kennicutt(1989), yields  gas mass fractions and stellar
truncation radii  in excellent agreement  with  observations (cf.  van
den Bosch 2000; van den Bosch \& Dalcanton 2000).

Despite these clear successes of the model, a closer inspection of the
model galaxies reveals two important  shortcomings.  The first problem
mainly   concerns LSB  disk  galaxies.    These  systems  form out  of
protogalaxies with   high angular  momentum   and  do  not produce   a
significant  bulge component.    This fact   that  LSB  galaxies  have
typically lower bulge-to-disk ratios than their HSB counterparts is in
good agreement with observations. However, the resulting stellar disks
of these LSB systems are  too  centrally concentrated. Although  their
surface density profiles are close to exponential at the outside, they
reveal a strong central  cusp in clear disagreement with observations.
The second problem  concerns the extent of the  gas disks.  The models
predict truncation radii of  the gas close  to the radii  $R_{\rm HI}$
where the HI column density has fallen to $\sim 10^{20} \cm^{-2}$.  In
real  disk  galaxies, however,  HI is observed  out to   radii well in
excess of $R_{\rm HI}$, again in clear contradiction with the models.

The problem with the  truncation radii seems straightforward to  solve
by  including viscosity, which will  transport disk  mass in the outer
parts to larger radii. However, at the same  time viscous transport is
oriented  inwards  at small radii. Thus,  whereas  viscosity seems the
obvious  solution for the problem  with  the truncation radii, it will
only  aggravate  the problem with the   central concentration of (LSB)
disks, and additional processes are required to  solve the problems at
hand. 

Including a simple model for a galactic wind  induced by SNe can expel
large amounts of baryonic matter from the disk,  but it does so with a
relative  efficiency  that is virtually independent   of radius.  This
implies that the actual density distribution is, except for an offset,
left intact. Additional energy input from AGNs, not taken into account
in our models,   is unlikely  to  solve  the problem with  the  overly
concentrated LSB galaxies,  since the mass  of massive  black holes is
strongly correlated  with  that  of the  bulges, which  are  virtually
absent in LSB galaxies.

Some studies in the past  have suggested that LSB galaxies differ from
their HSB  counterparts in  that they form  in density peaks  of lower
amplitude, rather than in peaks with more angular momentum (as assumed
here).  This implies that LSB galaxies form later and inside haloes of
lower  densities.   By  exploring  a  wide  range  of  MAHs  and  halo
concentrations we have shown that  even for this picture the LSB disks
are too centrally concentrated.

We  therefore  conclude that  understanding    the formation  of  disk
galaxies in CDM cosmologies faces  two important challenges.  We first
of  all need a   mechanism   that can   prevent the   angular momentum
catastrophe  which results  in disks being  an  order of magnitude too
small.  Furthermore, even if the    mass accretion is smooth and   the
angular momentum conserved, the disks that form, although of the right
size, are too centrally  concentrated.  The robustness of this problem
seems   to  suggest that  the     baryons  need an  angular   momentum
distribution  that is clearly distinct from  that of  the dark matter.
Similar results were recently  obtained by Navarro \& Steinmetz (2000)
and  van den Bosch, Burkert \&   Swaters (2001).  Navarro \& Steinmetz
used simple   scaling  relations  to show  that,    in  a $\Lambda$CDM
Universe, disk galaxies  only need to accrete  a small fraction of the
total baryonic   mass  to match  the   zero-point of  the Tully-Fisher
relation, but  must draw a  comparably  much  larger fraction  of  the
available  angular   momentum.  Van  den    Bosch, Burkert  \& Swaters
computed the angular momentum  distributions of  a sample of  low-mass
disk   galaxies from the  observed   rotation curves  and disk density
distributions.  A  comparison with the angular  momentum distributions
of  cold dark  matter haloes  found by  Bullock \etal (2000),  clearly
reveals  that disks lack  predominantly  low angular momentum material
compared to their dark matter haloes.

An interesting alternative to   considering a decoupling   between the
dark and baryonic mass components, is to change the nature of the dark
matter.   Recently, numerous  studies have  focussed  on  scenarios in
which the  dark matter  is warm  (WDM)  or self-interacting (SIDM). In
both cases  one  expects dark  matter haloes  to have constant density
cores with  less  substructure than in CDM  models  (e.g.,  Spergel \&
Steinhardt  2000;  Bode, Ostriker   \& Turok  2001).   This  not  only
alleviates  the problems with  disk rotation  curves  and the  angular
momentum catastrophe, but  it will also result in  disks that are less
centrally   concentrated.   In  order    to   investigate  whether   a
modification   of the nature  of  dark  matter might   solve the  disk
concentration problem, we have investigated models in which the haloes
have (large) constant density cores.  As expected, the resulting disks
are more  extended  as in  the CDM case,  and  with surface brightness
profiles that start to approach an  exponential form.  However, at the
same time  the  disks become  more self-gravitating, resulting  in the
formation of  relatively  massive bulges.   Therefore, the problem  of
producing bulge-less LSB  galaxies with exponential surface brightness
profiles  remains.  Before  concluding  that   the disk  concentration
problem persists  even   in WDM and SIDM  scenarios,   we caution that
throughout  we have made the assumption  that the  halo spin parameter
does not evolve with time.  Although  this seems a valid assumption to
make in the case of CDM, very little is known about the (evolution) of
the angular momentum distribution of haloes in alternative dark matter
scenarios. An investigation of the distribution of angular momentum in
haloes  of   warm  and/or self-interacting   dark   matter  will proof
extremely useful in this context.


\section*{Acknowledgments}

I am  indebted  to  Andreas  Burkert, St\'ephane  Charlot,   Guinevere
Kauffmann, Houjun Mo, and Simon White for stimulating discussions, and
to the anonymous referee for insightful comments.  Partial support for
this work was provided by the  National Science Foundation under Grant
No.  PHY94-07194,  and  by NASA  through Hubble  Fellowship Grant  No.
HF-01102.11-97.A awarded  by  the Space  Telescope Science  Institute,
which is operated by AURA for NASA under contract NAS 5-26555.


\end{document}